%% file: paper_ext.tex
\documentclass[conference, 10pt]{IEEEtran}
\IEEEoverridecommandlockouts
\usepackage{multirow}
\usepackage[table]{xcolor}

\usepackage{etoolbox}
\makeatletter
\patchcmd{\@makecaption}
  {\scshape}
  {}
  {}
  {}
\makeatletter
\patchcmd{\@makecaption}
  {\\}
  {.\ }
  {}
  {}
\makeatother
\def\tablename{Table}

\usepackage{verbatim} % allows for multi-line comments
\ifCLASSINFOpdf
   \usepackage[pdftex]{graphicx}
\else
\fi
\usepackage[cmex10]{amsmath}
\usepackage{amsfonts}
\usepackage{algorithmic}
\usepackage[caption=false,font=footnotesize]{subfig}
\usepackage{fixltx2e}
\usepackage{stfloats}
\usepackage{algorithm}

\let\labelindent\relax
\usepackage{enumitem}
\setlist[itemize]{leftmargin=*}

\usepackage{todonotes}

\begin{document}

%
% paper title
% can use linebreaks \\ within to get better formatting as desired
\title{Interactive Consistency in practical, mostly-asynchronous systems}

% author names and affiliations
% use a multiple column layout for up to two different
% affiliations

\author{
  \IEEEauthorblockN
  {
    Panos Diamantopoulos,
    Stathis Maneas,
    Christos Patsonakis, 
    Nikos Chondros and
    Mema Roussopoulos
  }
  \IEEEauthorblockA
  {
    Department of Informatics and Telecommunications\\
    National and Kapodistrian University of Athens \\
    Athens, Greece\\
    \{panosd,smaneas,c.patswnakis,n.chondros,mema\}@di.uoa.gr
  }
}

% conference papers do not typically use \thanks and this command
% is locked out in conference mode. If really needed, such as for
% the acknowledgment of grants, issue a \IEEEoverridecommandlockouts
% after \documentclass

% for over three affiliations, or if they all won't fit within the width
% of the page, use this alternative format:
% 
%\author{\IEEEauthorblockN{Michael Shell\IEEEauthorrefmark{1},
%Homer Simpson\IEEEauthorrefmark{2},
%James Kirk\IEEEauthorrefmark{3}, 
%Montgomery Scott\IEEEauthorrefmark{3} and
%Eldon Tyrell\IEEEauthorrefmark{4}}
%\IEEEauthorblockA{\IEEEauthorrefmark{1}School of Electrical and Computer Engineering\\
%Georgia Institute of Technology,
%Atlanta, Georgia 30332--0250\\ Email: see http://www.michaelshell.org/contact.html}
%\IEEEauthorblockA{\IEEEauthorrefmark{2}Twentieth Century Fox, Springfield, USA\\
%Email: homer@thesimpsons.com}
%\IEEEauthorblockA{\IEEEauthorrefmark{3}Starfleet Academy, San Francisco, California 96678-2391\\
%Telephone: (800) 555--1212, Fax: (888) 555--1212}
%\IEEEauthorblockA{\IEEEauthorrefmark{4}Tyrell Inc., 123 Replicant Street, Los Angeles, California 90210--4321}}

% use for special paper notices
%\IEEEspecialpapernotice{(Invited Paper)}

% make the title area
\maketitle

\begin{abstract}
\input{abstract}

\end{abstract}

\begin{IEEEkeywords}
Interactive consistency, Asynchronous, Consensus, Agreement, Byzantine

\end{IEEEkeywords}

% For peer review papers, you can put extra information on the cover
% page as needed:
% \ifCLASSOPTIONpeerreview
% \begin{center} \bfseries EDICS Category: 3-BBND \end{center}
% \fi
%
% For peerreview papers, this IEEEtran command inserts a page break and
% creates the second title. It will be ignored for other modes.
\IEEEpeerreviewmaketitle

\input{introduction}

\input{backgroundandrelatedwork}

\input{system_model}

\input{asyncic}

\input{implementation}

\input{evaluation}

\input{conclusion}

\bibliographystyle{IEEEtran}

%
%\bibitem{IEEEhowto:kopka}
%H.~Kopka and P.~W. Daly, \emph{A Guide to \LaTeX}, 3rd~ed.\hskip 1em plus
%  0.5em minus 0.4em\relax Harlow, England: Addison-Wesley, 1999.
%\newpage

\appendices
\input{icproof}

\input{appendix}

\end{document}

%% file: abstract.tex
Interactive consistency is the problem in which $n$ nodes, where up to 
$t$ may be byzantine, each with its own private value, run an algorithm 
that allows all non-faulty nodes to infer the values of each other node. 
This problem is relevant to critical applications that rely on the 
combination of the opinions of multiple peers to provide a service. 
Examples include monitoring a content source to prevent equivocation or 
to track variability in the content provided, and resolving divergent state 
amongst the nodes of a distributed system.

Previous works assume a fully synchronous system, where one can make strong 
assumptions such as negligible message delivery delays and/or detection of absent 
messages. However, practical, real-world systems 
are mostly asynchronous, i.e., they exhibit only some periods of synchrony during 
which message delivery is timely, thus requiring a different approach.
In this paper, we present a thorough study on \emph{practical interactive consistency}.
We leverage the vast prior work on broadcast and byzantine consensus algorithms to 
design, implement and evaluate a set of algorithms, with varying timing assumptions
and message complexity, that can be used to achieve interactive consistency in real-world
distributed systems.

% Most previous algorithms assume a synchronous system, where one can make 
% assumptions regarding message delivery delays and/or detection of absent 
% messages. Practical, real world systems are mostly asynchronous in nature, 
% thus requiring a different approach. In this paper, we present a thorough 
% study on \emph{practical interactive consistency}. We leverage 
% the vast prior work on broadcast and byzantine consensus algorithms to 
% design, implement and evaluate a suite of algorithms, each with its own 
% unique trade-off regarding processing overhead and message complexity, 
% that can be used to achieve interactive consistency in real distributed 
% systems. 

We provide a complete, open-source implementation of each proposed interactive 
consistency algorithm by building a multi-layered stack of protocols that 
include several broadcast protocols, as well as a binary and a multi-valued 
consensus protocol. Most of these protocols have never been implemented and evaluated in 
a real system before. We analyze the 
performance of our suite of algorithms experimentally by engaging in both 
single instance and multiple parallel instances of each alternative. 

%% file: introduction.tex
\section{Introduction}
\label{section:intro}

Interactive consistency (IC) is defined in a system of $n$ distinct nodes, each having its own private value, where up to $t$ may be byzantine (faulty). 
The goal is for all non-faulty nodes to compute the same vector of values.
For each non-faulty node, the corresponding slot in the vector should contain that node's private value.
%, where each honest node's slot contains its corresponding private value.

%na psaksw to paper tou xontrouli apo to eurosys me to chain replication pou ekane kai ta events
%isos einai sxetiko

\begin{comment}
This problem is closely related to applications that rely on the opinions of multiple witnesses to
provide a service. Our line of work revolves around the development of a distributed service that
will document the individual perspectives of multiple peers regarding the historic evolution
of a content source. For instance, this source could be the web site of some political party or 
even a government agency. In the past, there have been multiple cases in which content from these
sort of web sites has been either altered [[XXXadd references hereXXX]], or even completely removed in
some cases [[XXXadd references hereXXX]], according to the interests and/or the agenda of that time's
administration. Our system will be able to track and present these ``changes`` by combining the
opinions of multiple peers, each of which will belong in a separate administrative domain, thus
adding to its fault tolerance and credibility.
\end{comment}

%yes we do need to cite them here, we can just a couple for example : e.g. 'cite{lamport'}'
To date, related work regarding interactive consistency has provided
solutions tailored for synchronous systems (\cite{cloud.ic,thambidurai.ic,lincoln.ic,nasa.ic,lala.ic}).
These algorithms deliver useful theoretical insight and may be suitable in cases (e.g., shared memory multi-processor systems)
where one can make strong assumptions such as negligible message delivery delays (\cite{lamport.ic}) and/or 
detection of absent messages (\cite{cloud.ic,thambidurai.ic}). However, these assumptions are
ill-suited for practical, real-world distributed systems which,
in their vast majority are mostly asynchronous, i.e., they exhibit only some periods of synchrony during which message delivery is timely.

%There is a myriad of studies for asynchrnous environments that focus on two
In a fully asynchronous environment, researchers have proposed a myriad of algorithms for closely related topics, such as byzantine agreement and byzantine consensus. 
Consequently, one might assume interactive consistency can be easily achieved in an asynchronous setting, by a simple synthesis of one or more steps of these algorithms. 
However, this is not the case, as it is impossible to detect process crashes in a completely asynchronous system, where messages can take arbitrarily long to be delivered. 
Additionally, in an asynchronous system, it is impossible to guarantee simultaneously, that all honest parties inputs' are included in the computation (in our case, in the resulting vector of values), and that all honest parties are guaranteed to terminate, as proved in~\cite{zikas.impossible}. 
These are the reasons \emph{vector consensus} is considered the only achievable equivalent of interactive consistency for completely asynchronous systems (\cite{correia.multivalue.consensus}). 
In vector consensus, the only guarantee is that the resulting vector contains at least $2t+1$ values, of which at least $t+1$ were proposed by honest nodes.

Interactive consistency is required in a variety of real-world, critical applications.
%that rely on the reliable recording of the opinions of multiple witnesses. 
For example, a distributed fault-tolerant voting application runs a light-weight voting protocol during election hours~\cite{ddemos}.  At election end-time,  each node's local view of the cast votes may be inconsistent with the views of the others. 
Interactive consistency can be used here once, upon election end-time, to derive a single set of votes and produce the result.
As another example, an application may employ multiple peers to monitor the content delivered by a single source, as a means to verify its integrity.
This prevents the source from equivocating, i.e., distributing different content to different peers, with the additional benefit of being able to prove reliably if such equivocation took place. 
Another closely related application is the recording of the variability in web content, as a means to track censorship (\cite{censmon}), or other forms of personalization (\cite{PAN}). 
Other applications include the ability of sensors to reliably compute complicated functions that depend on the combination of inputs from other sensors (\cite{cbwsn}), 
%the reliable time stamping of events, as well as the handling of other non-deterministic values, for which there is no consistent approach in state machine replication (\cite{castro}), 
system diagnosis, such as failure detection and group membership, cloud computing (\cite{cloud.ic}), and other problems requiring global knowledge.
%Since we consider interactive consistency in a practical, mostly asynchronous setting, we do not focus on its application for performance-sensitive use cases, such as implementation of replicated state machines, or broadcast channels.

%Kalo einai i epomeni paragrafos na xoristei se 2 ''kommatia``. To ena kommati na perigrafei
%ta theoritika challenges ta opoia prepei na antimetopisoume ta opoia einai sti fisi ton
%asigxronwn algorithmwn kai den exoun na kanoun me kapoia sigkekrimeni ilopoihsh. Auta ta
%impossibility results einai to FLP kai tou Zika to input completeness kai termination.
%Ta practical challenges prokiptoun apo unrealistic ipotheseis pou kanoun theoritika papers
%kai ta briskoume mprosta mas otan pame na ilopoihsoume autous tous algorithmous, kai einai
%aneksartita apo to an o algorithmos einai gia sigxrono i gia asigxrono. Auta einai to infinite
%node memory, a lossless network (omg), message corruption (?)

In this paper, we present algorithms for solving interactive consistency
in real-world distributed systems, with the minimal possible timing assumptions.
We leverage prior work on broadcast and byzantine consensus protocols to design our algorithms. 
% Since byzantine fault-tolerant
% algorithms are generally known to exhibit high message complexities (e.g.,~\cite{bracha.async.consensus} has a message complexity of $O(n^3)$),
% we design our algorithms to tradeoff higher computational overhead for lower message complexity depending on the intended 
% environment of use.  For example, some of our algorithms rely on digital signatures and require
% less communication. In cases where network communication is the bottleneck, such as wide area networks (WANs)
% and the Internet, it may be more efficient to verify a few signatures than to suffer increased message complexity.
 We describe how we directly address, or circumvent, both theoretical and practical challenges that arise in solving IC. 
Examples of theoretical challenges are the well-known FLP impossibility result (\cite{FLP}) and the impossibility of simultaneously achieving input completeness and guaranteed termination (\cite{zikas.impossible}). 
Practical challenges, on the other hand, are a result of assumptions that several theoretical papers use to prove their algorithms, but that, unfortunately, do not hold in practice.
Examples are unbounded memory at each node and a loss-less and/or corruption-free network (\cite{lala.ic,bracha.async.consensus}).
Moreover, we have formally proved the correctness of our proposed algorithms.
% due to space limitations, we refer the reader to the Appendix of the extended version of this paper (\cite{dsgPic}).

% Other algorithms in our suite rely on the more efficient scheme of message authentication codes (MACs). They do, however,
% require bigger messages to be exchanged, which makes them more suitable for environments where
% processing power is limited.
To evaluate our algorithms, we first analytically compare their message complexity.
We implement all of the proposed algorithms and present experimental measurements
that consist of both single and multiple parallel instances. We compare the algorithms'
performance in terms of throughput and latency and draw conclusions as to the appropriateness
of using each algorithm in varied network environments.

%the trade-off between the use (or not) of digital signatures and message
%complexity in interactive consistency algorithms, as well as the efficiency of each alternative,
%we implemented all of the proposed algorithms and engage in a series of experiments. These involve
%both serial and parallel executions of each proposed algorithm to evaluate their performance
%in terms of throughput and latency.[[XXXthis paragraph is incomplete, we need to add something regarding results hereXXX]]
In summary, we make the following contributions:
\begin{itemize}
 \item We present a study of interactive consistency in \emph{practical, real-world systems}, illustrate the theoretical and practical challenges that need to be addressed, and propose algorithms for achieving interactive consistency in such environments.
%  \item We propose a suite of algorithms that can be used to achieve interactive consistency. We compare the algorithms
%  analytically and show that each alternative provides its own unique trade-off regarding timing assumptions and message complexity.
 \item We provide an open-source implementation of each of the proposed alternatives. This required the development
 of a protocol stack that includes various asynchronous broadcast primitives, as well as a binary and a multi-valued byzantine consensus protocol. Some of these protocols (e.g., \cite{correia.multivalue.consensus,bracha.async.consensus}),
 have never been implemented and evaluated (to our knowledge) in a real system.
 \item We evaluate our algorithms experimentally by running both serial and parallel executions of
 each algorithm and compare their performance in terms of throughput and latency, in both LAN and WAN settings. We find 
 that simple protocol variations that restrict the behavior of malicious nodes improve performance.
\end{itemize} 

% The remainder of this paper is organized as follows. In Section~\ref{section:backgandrw}, we 
% present background information needed for the non-expert reader and prior related works.
% In Section~\ref{section:sysmodel}, we describe our system model.   
% In Section~\ref{section:asyncic}, we describe in detail the suite of algorithms we design to solve
% interactive consistency in practical, mostly asynchronous environments. In Section~\ref{section:implementation}, we describe the implementation
% of our protocol stack and the prototypes of each of our algorithms.
% In Section~\ref{section:evaluation}, we present results from an
% experimental performance comparison of our algorithms and in Section~\ref{section:conclusion}, we conclude.

%% file: backgroundandrelatedwork.tex
\section{Background and Related Work}
\label{section:backgandrw}

Our study has unveiled a large incoherence in the bibliography,
regarding the terms ``Byzantine Agreement'' and ``Byzantine Consensus''. These are often used to refer to the same problem 
(e.g.,~\cite{FLP} and~\cite{toueg.trustdealer.consensus}), while others, e.g.,~\cite{patra.byz.agree.mv},
use the terms interchangeably, even though these are two distinct problems. 
There is also inconsistent use of the term ``Interactive Consistency'', e.g.,~\cite{milosevic.weakic,postma.ic}.
To alleviate any confusion and for clarity, we start with some basic definitions.

{\bf Byzantine Agreement.} Assume a system of $n$ nodes, where a single source $n_i$
has a private value $v_i$, and the following must be achieved:

\begin{itemize}
 \item \emph{Agreement:} All non-faulty nodes must agree on the same value.
 \item \emph{Validity:} If $n_i$ is non-faulty, then the agreed upon value by all non-faulty
 nodes is $v_i$.
 \item \emph{Termination:} All non-faulty nodes must eventually decide on a value. 
\end{itemize} 

This problem, also known as the ``Byzantine Generals Problem'', was introduced by
Lamport et al.~\cite{lamport.byz.gen}. 
Earlier work has proved there is no
solution for the asynchronous case~\cite{bracha.async.consensus}, when the source
is faulty.  
%even in the
%presence of a single fault, since $p$ may fail to send any messages. Therefore,
%it is a necessity for $p$ to be non-faulty. 
Agreement algorithms that tolerate
byzantine failures of (non-source) nodes in asynchronous systems are presented in~\cite{bracha.byz.agree}
and~\cite{rabin.byz.agree}.

{\bf Byzantine Consensus.} Assume a system of $n$ nodes, where each node $n_i$
has a private value $v_i$, and the following must be achieved:

\begin{itemize}
 \item \emph{Agreement:} All non-faulty nodes must agree on the same value $v \in \{v_1,...,v_n\}$.
 \item \emph{Validity:} If all non-faulty nodes have the same initial value
 $v$, then the agreed upon value by all non-faulty nodes is $v$.
 \item \emph{Termination:} All non-faulty nodes must eventually decide on a value.
\end{itemize}

The byzantine consensus problem is one of the most studied topics in distributed systems
and the main topic of the well-known FLP impossibility result (\cite{FLP}). There are
several types of consensus protocols. The first distinction revolves around determinism
(or non-determinism). In a deterministic consensus protocol, given the set of input
values on all nodes, the message schedule and the failures that occur (if any), the
result will always be the same. Deterministic consensus protocols
require a synchronous system (\cite{dolev.pinakaki}). In a purely asynchronous system, consensus
can be achieved by randomization.
FLP is circumvented by having nodes locally toss a coin to decide on their input values, in round $r+1$,
in cases where consensus cannot be achieved in round $r$. Thus, the result may
be different across executions with the same inputs.
Examples of randomized protocols that employ the local coin construct are
introduced by Bracha~\cite{bracha.async.consensus}, Bracha and Toueg~\cite{bracha.toueg.consensus} and
Ben-Or~\cite{benor.consensus}. These algorithms guarantee eventual termination after a probabilistic number
of rounds. In~\cite{toueg.trustdealer.consensus}, a trusted, non-faulty dealer is additionally employed to bound 
the number of rounds required to achieve consensus. 
%Its function is to, initially, compute a random bit sequence (secret) and then to
%reliably distribute shares of that secret to all of the $n$ nodes in the system. 
In our work, we leverage the randomized approach by Bracha to ensure termination because we believe
it is controversial to assume a trusted entity in an otherwise byzantine environment.

Other works (\cite{canetti.consensus},~\cite{patra.consensus}) leverage verifiable secret sharing techniques to implement a shared, or, common
coin. These consensus algorithms are polynomially efficient and terminate in a constant number of rounds.
%The common coin construct, with probability $\frac{1}{4}$, delivers the same value $\sigma \in \{0,1\}$ to all non-faulty nodes, provided that up 
%to $t$ nodes are faulty. 
Canetti et al.~\cite{canetti.consensus} present one of most well-established and signature-free common coin protocols.
However, this protocol, although polynomial, is complex to implement and has very high bit complexity~\cite{mostefaoui.consensus}.
Most{\'{e}}faoui et al.~\cite{mostefaoui.consensus} employ the common coin protocol that is presented in~\cite{cachin.coin} which has guaranteed termination but requires a trusted dealer.
We did not consider these algorithms as they are either inefficient or require a trusted dealer.

One last distinction, regarding consensus protocols, revolves around the agreed upon value.
All of the aforementioned protocols are binary consensus protocols, i.e., the agreed upon value
is $v \in \{0,1\}$. In the multi-valued consensus protocol of Correia et al.~\cite{correia.multivalue.consensus},
the set of values $V$ is of arbitrary size. In our work, when needed, we achieve multi-valued consensus
by using primitives such as reliable broadcast (described later) and binary consensus.

{\bf Failure Detectors.} In~\cite{chandra.failure}, Chandra and Toueg proposed a solution 
for the consensus problem, in an asynchronous crash-fault environment, introducing a module 
called \emph{failure detector} (FD).
%Each process has access to a local failure detector module, which 
%holds a list with all process it considers crashed.Failure detectors are defined in terms of 
%completeness and accuracy.
%\begin{itemize}
%\item \emph{Strong completeness}: Eventually every process that crashes is permanently suspected 
%by every correct process.
%\item \emph{Weak completeness}: Eventually every process that crashes is permanently suspected 
%by some correct process. 
%\item \emph{Strong Accuracy}: No process is suspected before it crashes. 
%\item \emph{Weak Accuracy}: Some correct process is never suspected. 
%\item \emph{Eventual Strong Accuracy}: There is a time after which, correct nodes are not 
%suspected by any correct process. 
%\item \emph{Eventual Weak Accuracy}: There is a time after which, some correct process is 
%never suspected by any correct process.
%\end{itemize}
There is extensive bibliography that expands the family of FDs to 
a number of applications (\cite{doudou.muteness, guerraoui.weakest, helary.global, kihlstrom.byzantine}).
% The muteness failure detector presented in~\cite{doudou.muteness} is a
% failure detector can detect nodes that, after a time, either crash, or behave arbitrarily.
% In~\cite{guerraoui.weakest}, FDs are used to solve the transaction commit problem in distributed databases. 
% Mostefaoui et al.~\cite{helary.global}, use perfect failure detectors to compute global data, used in the distributed 
% termination detection problem. Finally, Kihlstrom et al.~\cite{kihlstrom.byzantine}, extend 
% the work of Chandra and Toueg and propose a failure detector that can 
% effectively detect Byzantine behaviour.
%Despite their extensive use in theoretical models, the implementation and 
%use of FDs remains infeasible in real-world distributed systems, 
%as their assumptions cannot be met. 
In~\cite{chandra.weakest}, 
Chandra and Toueg define the weakest FD capable of solving consensus in asynchronous crash fault 
systems. However, this failure detector requires known bounds on node processing speed 
and message delivery, that hold after a Global Stabilization Time (\cite{larrea.optimal}). The same assumptions hold for the Byzantine Failure detector 
introduced in \cite{kihlstrom.byzantine}. 
However, these assumptions are unlikely to hold in real-world distributed systems, rendering both FDs unimplementable 
(\cite{garg.implementable, correia.consensus.survey}). 
%To the best of our knowledge, there is no failure 
%detector that requires weaker synchrony assumptions and can detect byzantine faults.

{\bf Broadcast Primitives.} 
All asynchronous consensus algorithms employ some form of reliable broadcast protocol, 
where a source \emph{broadcasts} a message $m$, and every correct node eventually \emph{delivers} $m$ (e.g., via an up-call to the application).  
Such a broadcast satisfies the following properties (\cite{hadzilacos.modular}):
\begin{itemize}
 \item \emph{Validity}: If a non-faulty node broadcasts a message $m$, then it
 eventually delivers $m$.
 \item \emph{Agreement}: If a non-faulty node delivers a message $m$, then all 
 non-faulty nodes eventually deliver $m$.
% This is often described as two individual properties (\cite{cachin.secure}):
%   \begin{itemize}
%    \item \emph{Consistency}: If some non-faulty node delivers a message $m$ with an $ID$ and another
%     non-faulty node delivers a message $m'$ with the same $ID$, then $m=m'$.
%     \item \emph{Totality}: If some non-faulty node delivers a message $m$ with an $ID$, then all non-faulty 
%     nodes will eventually deliver some message with the same $ID$.
%   \end{itemize}
 \item \emph{Integrity}: For any message $m$, every non-faulty node delivers $m$
 at most once \emph{iff} $m$ was previously broadcast by $sender(m)$.
\end{itemize}

%As stated in \cite{cachin.secure}, reliable broadcast is equivalent to the Byzantine Generals Problem in the presence 
%of arbitrary faults. 
In~\cite{bracha.async.consensus}, Bracha introduced a $\frac{n}{3}$-resilient reliable broadcast 
primitive (RBB, for Reliable Broadcast of Bracha) to solve the consensus problem.
Another type of broadcast primitive, with lower message complexity, is \emph{consistent broadcast} (CB).
CB is designed to relax the \emph{agreement} property of reliable broadcast, by allowing \emph{some} non-faulty nodes to deliver $m$, while others may deliver nothing. 
The standard implementation of consistent broadcast is \emph{Reiter's echo multicast}~\cite{reiter.secure}.  

% Another type of  broadcast primitive, with lower message complexity, is \emph{consistent broadcast} (CB), which is 
% designed not to satisfy the \emph{totality} property that is often too expensive. Consistent broadcast replaces 
% reliable broadcast in applications where \emph{totality} can be ensured by other means. The standard implementation
% of consistent broadcast is \emph{Reiter's echo multicast}~\cite{reiter.secure}.  

{\bf Interactive Consistency.} Assume a system of $n$ nodes, where each node $n_i$
has a private value $v_i$, and the following must be achieved:

\begin{itemize}
 \item \emph{Agreement:} All non-faulty nodes must agree on the same vector of values
 $V=[v_1,...,v_n]$.
 \item \emph{Validity:} If the private value of the non-faulty node $n_i$ is $v_i$, then
 all non-faulty nodes agree on $V[i]=v_i$.
%  If $n_i$ is faulty, then all non-faulty nodes can agree on any value for $V[i]$.
 \item \emph{Termination:} All non-faulty nodes must eventually decide on a vector $V$.
\end{itemize}

%Theloume na poume oti eimaste practical guys kai oti dinoume ena sinolo apo solutions
%gia to problima auto se asigxrono sistima kai oti episis kanoume evaluate ola ta diaforetika
%approaches. Kai oti eimaste oi protoi pou to kanoume se asigxrono sistima, me kapoies paradoxes
\begin{comment}
Interactive consistency (IC) was first introduced and studied by Lamport et al. \cite{lamport.ic}.
Since, it has been the topic of several research papers (\cite{cloud.ic}, \cite{thambidurai.ic}, \cite{lincoln.ic},
\cite{nasa.ic}, \cite{lala.ic}). This line of work proposes algorithms
for synchronous systems where one can assume negligible, or bounded, message delays
(\cite{thambidurai.ic}, \cite{nasa.ic}, \cite{lamport.ic}), or even detect the absence of
messages (\cite{cloud.ic}, \cite{lincoln.ic}). While these approaches might be feasible in
environments such as shared memory multi-processors or digital flight control systems, we
believe that they are ill-suited for practical, real-world distributed systems which are
mainly asynchronous in nature. To our knowledge, and to date, interactive consistency has
not been studied in an asynchronous setting (L. Lamport, private communication).
Thus, we consider our work as the first attempt of solving this problem in a practical,
real world setting.
\end{comment}
Interactive consistency was first introduced and studied by Pease et al.~\cite{lamport.ic}, and 
has been the topic of several research papers (\cite{cloud.ic,thambidurai.ic,lincoln.ic,nasa.ic,lala.ic,benor.ic}),
focusing on synchronous systems. 
%where one can assume negligible, or bounded, message delays
%(\cite{thambidurai.ic,nasa.ic,lamport.ic,benor.ic}), or even detect the absence of
%messages (\cite{cloud.ic,lincoln.ic}). 
While these approaches might be feasible in
environments such as shared memory multi-processors or digital flight control systems, we
believe they are ill-suited for practical, real-world distributed systems.
In~\cite{milosevic.weakic} and~\cite{postma.ic}, the authors provide solutions to various forms of consensus, despite their title references to IC.

A closely related problem to IC is vector consensus. These two problems differ only in terms of the \emph{Validity}
condition. Vector consensus delivers a vector with at least $2t+1$ values, where at least
$t+1$ values were proposed by non-faulty nodes. The reason for this difference is that in asynchronous
systems, it is impossible to ensure that the resulting vector has the proposals of all non-faulty
nodes (\cite{zikas.impossible,correia.multivalue.consensus}).
%, since they can be arbitrarily delayed 
%Several papers have studied this problem in both synchronous and asynchronous settings
%(\cite{correia.multivalue.consensus}, \cite{doudou.muteness}, \cite{benor.ic}, \cite{vaidya.vectorconsensus}, \cite{correia.vectorconsensus}, and~\cite{canetti.coreset} where it was named ``agreement on a core set''). 
%Their algorithms use a variety of techniques such as wormholes (\cite{correia.vectorconsensus}),
%failure detectors (\cite{doudou.muteness}), and leveraging the equivalence of multi-valued consensus
%and vector consensus (\cite{correia.consensus.survey}).

%% file: system_model.tex
\section{System Model}
\label{section:sysmodel}

We assume a distributed system consisting of $n$ nodes that are fully connected over a network. 
The network is mostly asynchronous, i.e., it exhibits one (or more, depending on the algorithm) period of synchrony, during which message delivery is timely. 
%A synchronous period is a period of time at the end of which the messages of all non-faulty nodes are successfully delivered.
%(Section \ref{section:asyncic}).
%After this synchronous round is completed the network may behave completely asynchronous and no assumptions can be made regarding the time messages are delivered. 
The network can drop, delay, duplicate, or deliver messages out of order. 
However, we assume that messages are eventually delivered, provided that the corresponding senders keep on retransmitting them.
We assume authenticated channels, where the receiver of a message can always identify its sender.
Each node has a public/private key pair and all nodes know the others' public keys.
We use these keys to implement authenticated channels, and sign messages where needed.

We assume a Byzantine failure model where nodes may deviate
arbitrarily from the protocol. We allow for a strong adversary that
can coordinate faulty nodes. However, we assume he cannot delay
the delivery of messages, or processing on correct nodes beyond the
system's synchrony assumptions. The adversary is also assumed to be
computationally bounded, meaning he cannot subvert common cryptographic
techniques such as signatures and message authentication codes (MACs).

%We assume an asynchronous distributed system comprised of $n$
%nodes that are fully connected over a network. Each
%node has a public/private key pair and all nodes know the
%others' public keys.

%Nodes communicate by exchanging messages. We assume that
%the receiver of a message can always identify its sender. This is
%commonly achieved by employing authenticated channels, i.e., using
%asymmetric cryptography to establish a shared symmetric key that is
%used to secure pair-wise communication. The network
%can drop, delay, duplicate, or deliver messages out of order. However,
%we assume that messages are eventually delivered, provided that the
%corresponding senders keep on retransmitting them.

% We assume a Byzantine failure model where nodes may deviate
% arbitrarily from the protocol. We allow for a strong adversary that
% can coordinate faulty nodes. However, we assume he cannot delay
% the delivery of messages, or processing on correct nodes beyond the
% system's synchrony assumptions. The adversary is also assumed to be
% computationally bounded, meaning he cannot subvert common cryptographic
% techniques such as signatures and message authentication codes (MACs). 

%% file: asyncic.tex
\section{Practical Interactive Consistency}
\label{section:asyncic}

\subsection{Adapting approaches from synchronous systems}
\label{subsection:fromsynchronous}
The original algorithm of Pease et al.~\cite{lamport.ic} requires a total of $t+1$
rounds to achieve IC in a synchronous system, tolerating up to $t$ faults, with a total message complexity of
$(t+1)n^2$. Our first approach is to adapt the same algorithm
by simulating synchronous rounds with timeouts.
Messages delivered after the time frame of each round, will be disregarded and counted towards the $t$ system faults, according to the model presented in~\cite{dolev.faults}. 

Two issues arise from the use of timeouts, as highlighted in~\cite{guerraoui1997consensus}. The first one is efficiency. 
Assuming a timeout value of $T_r$ for each round, the system will always require a constant amount
of time, i.e., $(t+1){T_r}$, to execute a request even in the presence of a single failure.
The second is choosing a correct value for $T_r$. If we choose a conservative approach and
set a large value for $T_r$, we could increase the execution time of the algorithm
dramatically, thus, making it less practical. On the contrary, a small value might
cause some slow nodes, who are otherwise correct, to be considered faulty. If this
occurs multiple times, as is the case when one relies on multiple timeouts, it is
possible that we will exceed the upper bound $t$ of total failures in the system.

To avoid the issues associated with multiple timeouts, one might attempt to reduce IC to
Byzantine Agreement (BA), by running $n$ parallel instances of BA, as it was suggested for synchronous
systems (\cite{fischer1983consensus}). In each instance, a node $n_i$ would spread its private value
$v_i$ to the rest of the system. In a synchronous setting, this would result in all non-faulty nodes having
the same vector of values. However, in a completely asynchronous environment, BA is impossible (\cite{bracha.async.consensus}),
as a crashed node may never even start its instance of BA, and nodes cannot distinguish between crashed nodes and slow nodes.
Therefore, the non-faulty nodes need to decide, at a certain point, to exclude the suspected crashed nodes from IC and store a default (e.g., \emph{null})
value at the slot corresponding to each crashed node. Thus, they need a synchronization point, where they decide on the result vector; we call this point a \emph{barrier}. 
This synchrony assumption allows for the circumvention of the impossibility of simultaneously achieving input completeness
and guaranteed termination in a purely asynchronous system (\cite{zikas.impossible}).

% To avoid the issues associated with multiple timeouts, one could attempt to 
% leverage the vast prior work on {\bf asynchronous} Byzantine Agreement (BA) and consensus
% algorithms. A simple solution to IC would be to run $n$ parallel instances of BA,
% where each node $n_i$ would spread its private value $v_i$ to the rest of the system. 
% In a synchronous setting, this would result in all non-faulty nodes having the same vector of values.
% However, this approach is inadequate in a completely asynchronous environment, as a crashed node may
% never even start its instance of BA, and nodes cannot distinguish between crashed nodes and slow nodes.
% Therefore, in the asynchronous setting, the non-faulty nodes need to decide, at a certain point, to exclude the suspected crashed nodes from IC and store a default (e.g., \emph{null}) value at the slot corresponding to each crashed node. 
% Thus, they need a synchronization point, where they decide on the result vector; we call this point a \emph{barrier}. 
% This synchrony assumption allows for the circumvention of the impossibility of simultaneously achieving input completeness and guaranteed termination in a purely asynchronous system (\cite{zikas.impossible}).

The introduction of the barrier introduces a new challenge as, at that point, a BA instance may have \emph{delivered} 
the result in some nodes but not yet in others. This, for example, may be triggered by an adversary starting his own BA 
instance near the barrier. Thus, honest nodes will need to achieve consensus, for each individual slot of the result 
vector, on the value to be placed in that slot. We observe that the barrier splits the procedure in two phases. 
We call the first phase the \emph{value dissemination phase}, where we assume the network delivers all messages of non-faulty nodes by the end of the phase. 
Recall that, as we stated in the first paragraph of this section, messages delivered after the time frame of the first phase will be counted towards the $t$ system faults.
We call the second phase, the \emph{result consensus phase}, which can be completely asynchronous.
Note that we have employed the costly BA approach for the first phase, but have shown that a consensus phase is still required.

\subsection{Solution using Multi-Valued Consensus}
\label{subsection:avoidba}
With these observations, we seek less costly alternatives for the first phase, i.e., avoiding BA. Our first approach is to use a simple point-to-point message exchange, where each node announces its own private value to the rest of the system. As this exchange is unrestricted, it may result in each honest node receiving a different value from a malicious node. Thus, during the result consensus phase, nodes need to agree on the value to be placed in each slot of the result vector.
We employ the multi-valued consensus (MC) algorithm from~\cite{correia.multivalue.consensus}; recall that this algorithm utilizes a binary consensus and a reliable broadcast primitive. We want to refrain from making any further synchrony assumptions, thus, making the result consensus phase completely asynchronous. In order to circumvent FLP, which states that achieving deterministic consensus is impossible in purely asynchronous systems, we employ a randomized consensus protocol. We use Bracha's binary consensus (BC) and reliable broadcast (RBB) primitives from~\cite{bracha.async.consensusfirstpaper}, and we run $n$ parallel instances of MC, one for each value of the vector.

This algorithm (\emph{IC,MC-RBB}) achieves IC because, regardless of the unrestricted value dissemination phase, each instance of MC ensures that nodes agree on a single value for each slot of the result vector respectively.
\emph{(IC,MC-RBB)} uses only one synchrony barrier, as opposed to the adaptation of Pease's algorithm which needs $t+1$.
Its overall message complexity is $10n^4 + 5n^3 + n^2$. For full derivation details of all message complexities, and formal proofs of correctness of all IC algorithms, we refer the reader to the Appendix.
% of the extended version of this paper (\cite{dsgPic}).
% Figure~\ref{figure:mc_rbb} demonstrates the message exchanges for this protocol.
% \begin{figure}[ht!]
% \centering
%   \includegraphics[width=0.47\textwidth, scale=0.7]{ic_mc_rbb}
% \caption{Diagram of message exchange for \emph{IC,MC-RBB}, for a single value of the result vector (repeated $n$ times to achieve IC).}
% \label{figure:mc_rbb}
% \end{figure}

\subsection{Solution using Binary Consensus}
\label{subsection:icbcrbb}
\begin{figure}[ht!]
\centering
  \includegraphics[width=0.47\textwidth, scale=0.7]{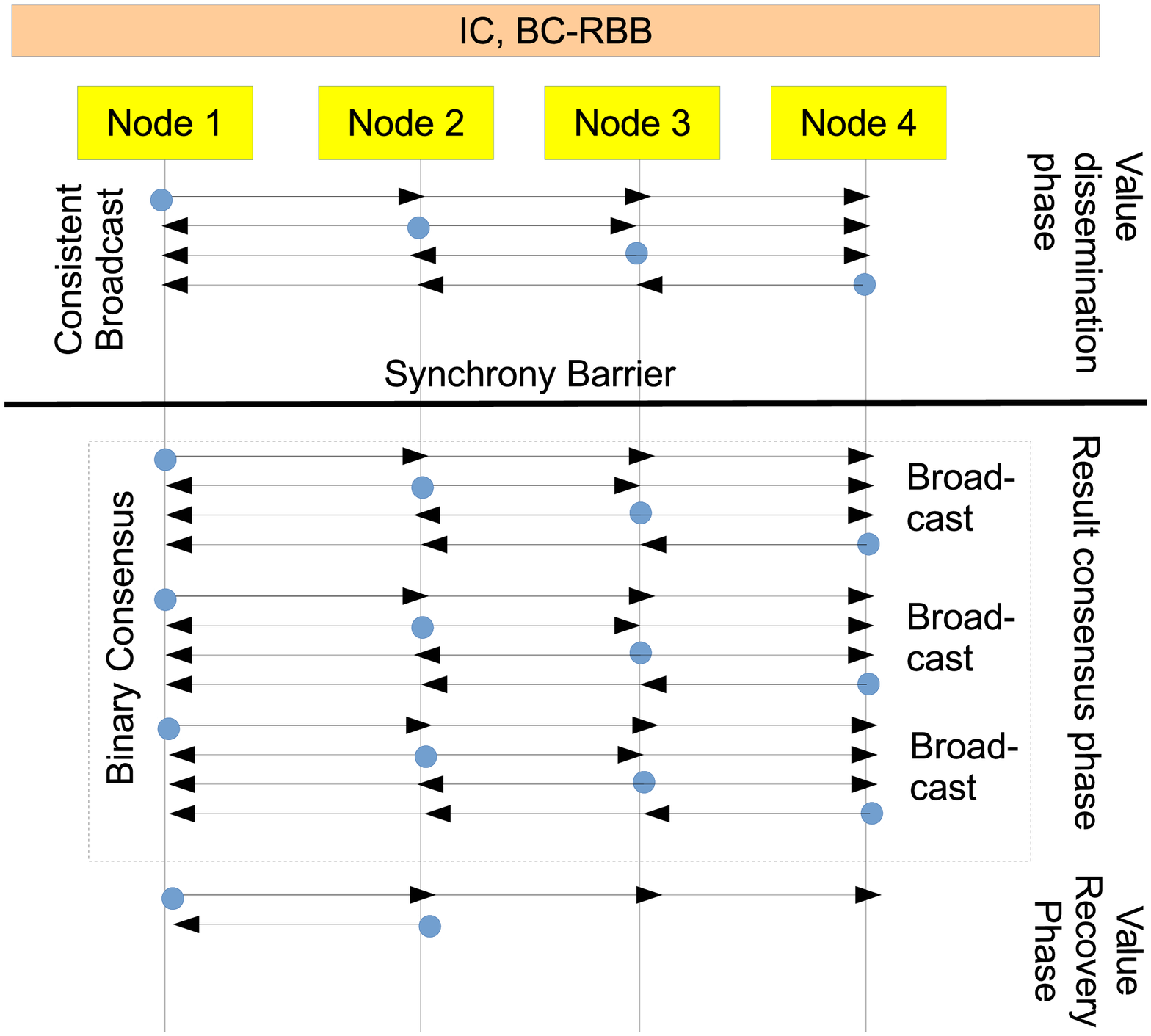}
\caption{Diagram of message exchanges for \emph{(IC,BC-RBB)}, for a single value of the result vector (repeated $n$ times to achieve IC).}
\label{figure:bc_rbb}
\end{figure}

Our next approach reduces the aforementioned message complexity. We observe multi-valued consensus uses one binary consensus and two reliable broadcast instances.
We avoid the use of MC by changing the subject on which consensus is required. In the previous algorithm, the consensus question is ``what is the actual value to be placed in the corresponding slot of the result vector?'', because the first (value dissemination) phase is insecure. 
We make the first phase secure by using Consistent Broadcast (CB, \cite{cachin2001secure}).
Here, the source $n_i$ first sends its value $v_i$ to each node; then it collects signed endorsement responses. 
A recipient node endorses only the first value for each broadcast. 
Once $n-t$ such responses are accumulated, the sender forms a uniqueness certificate $c_i$ that includes these endorsements, and sends \textless$n_i,c_i$\textgreater\text{ }to the rest of the nodes. CB \emph{delivers} $v_i$ \emph{iff} $c_i$ has at least $n-t$ valid signatures.
Assuming signatures are unforgeable, it is impossible for a malicious node to construct two valid certificates for two different values.
Thus, this protocol bounds the sender to either send a single value, or not send a value at all. 
As this value is guaranteed to be unique, we change the question of the result consensus phase to ``is there a value to be placed in the corresponding slot of the result vector?''. This question can now be answered by a binary consensus protocol, and we utilize Bracha's protocol (\cite{bracha.async.consensusfirstpaper}) in our approach.
Figure~\ref{figure:bc_rbb} depicts message exchanges for this protocol.

An outcome of $0$ from BC causes each node to place the \emph{null} value in the corresponding slot of the result vector. 
Accordingly, a result of $1$ from BC instructs each node to place the (unique) value $v_i$ in the result vector.
There are cases, however, where a consensus instance may produce a result different than the opinion with which an honest node entered BC.
This can happen when the corresponding instance of CB \emph{delivered} the value $v_i$ at some nodes, but not at others 
(e.g., when a malicious CB source sends the value, along with the uniqueness certificate, only to some nodes).
Thus, a node may possess a value for this slot, and the result of consensus may be $0$, in which case it simply replaces the value with \emph{null}. 
However, the contrary may also happen, where a node did not possess a value when it entered BC, but consensus resulted in $1$. 
For this case, we add a final \emph{recovery} phase, where a node asks all other nodes for the correct value of the $i^{th}$ position of the result vector. Any node that receives such a message replies with the \textless$v_i,c_i$\textgreater\text{ }tuple it possesses.
At least one honest node is guaranteed to exist and submit such a reply; this is because, by definition of BC, if all honest nodes entered consensus with $0$, the result would have been $0$. As the result is $1$, at least one honest node exists which has entered consensus with $1$, thus possessing the correct value and uniqueness certificate for it.

To summarize, this IC algorithm \emph{(IC,BC-RBB)} achieves IC because: a) during the value dissemination phase, an honest node either obtains a value guaranteed to be unique, or no value at all, b) during the result consensus phase, all nodes agree, for each slot of the result vector, whether to place a (guaranteed unique) value, or the \emph{null} value, and c) during the recovery phase, any honest node is guaranteed to obtain missing values. The overall complexity of \emph{(IC,BC-RBB)} is $6n^4 + 3n^3 + 3n^2$ messages and $n^3+2n^2$ signature operations. We formally prove the correctness of \emph{(IC,BC-RBB)} in Appendix~\ref{subsection:icbcrbbproof}.

\input{eventual}

%% file: eventual.tex
\subsection{Eventual Interactive Consistency}
\label{subsection:eventual}
So far, we present solutions to IC using one or more synchrony barriers, as the problem is unsolvable in a completely asynchronous setting.
There is, however, a weaker version of the problem, which can be solved without timing assumptions (synchrony barrier), which we introduce and briefly outline two solutions.
We call this weaker version \emph{Eventual Interactive Consistency} (EIC). In EIC, the \emph{Agreement} part of the problem's definition is as follows:
\begin{itemize}
 \item \emph{Agreement:} All non-faulty nodes must \bf{eventually} agree on the same vector of values
 $V=[v_1,...,v_n]$.
\end{itemize}

In this scheme, a non-faulty node will eventually build the result vector, containing all private values from all non-faulty nodes. Until it does, however, it may have empty slots for values it does not yet know about. In practice, the result vector will be built slot by slot, and instead of a single up-call to deliver the complete vector, multiple up-calls will take place. Each up-call will inform the application the vector was augmented by one more value. This version of the problem is suitable, for example, for applications which gather opinions. The idea is, the application can serve the already gathered opinions to clients, with empty slots for the currently unknown ones, either immediately upon request, or when a system-defined threshold of entries has been filled in the vector. Eventually, when all non-faulty nodes provide their opinions, empty slots in the result vector will represent failed nodes. 
If the vector is used before it is completed, the only guarantee provided is that on a subsequent access, the previous entries in the vector will still be included, potentially along with newer ones.

One simple approach to achieve this is by using a version of Reliable Broadcast (RB). All nodes start one instance and, by definition of RB, eventually all correct nodes' broadcasts deliver the intended value. When one of these RB instances completes, the corresponding slot of the vector will be filled and a new notification will be sent to the upper level application. This approach leaves management of the result vector completely up to the application.

A more involved approach which, however, preserves and manages the result vector as well, is the use of a byzantine fault tolerant Replicated State Machine (RSM), such as~\cite{castro,bessani2014state} enhanced to handle byzantine clients (\cite{clement2009making,liskov2005byzantine}). 
Each node of EIC becomes both a replica and a client of the same RSM. Each IC node, as a client of the RSM, posts its private value to the RSM as a \emph{write} operation. The application running on top of the RSM receives this \emph{write} operation and accepts it only when no prior value is known for the sending node and this instance of EIC. Whenever an external client requests the EIC result vector, it is dynamically compiled from entries already posted from nodes. A malicious node cannot harm the system as long as the RSM's fault tolerance level (typically $t < \lceil\frac{n}{3}\rceil$) is not breached, while as a client, it is prohibited from posting more than one value by the aforementioned functionality running at the application layer behind the RSM.

%% file: implementation.tex
\section{Implementation}
\label{section:implementation}

We developed an open-source protocol in Java to implement and evaluate our suite of algorithms. At the foundation lies an authenticated
channels layer that uses SSL and manages message passing and timeout events; SSL provides for authentication and message integrity.
We also provide alternatives for direct TCP/IP communication (without strong authentication), as well as, an intra-process communication infrastructure that allows us to run our unit tests and verify the correctness of our implementation.
The remaining layers are agnostic of the network layout or communication means, as they simply register event handlers to process
incoming messages. Finally, we simulate loss-less channels by creating one output queue for each node, where each
queue is handled by a different thread. A message is deleted from a queue only when the sender
receives an acknowledgment for that specific message by the destination.

On top of this foundation, we implement Consistent Broadcast, and
the signature-free Reliable Broadcast primitive of Bracha (RBB).
We then implement Bracha's binary consensus (BC) protocol (\cite{bracha.async.consensus}),
which uses RBB. Finally, we implement the multi-valued consensus protocol (MC) of Correia
et al.~\cite{correia.multivalue.consensus}, using BC and RBB.

\begin{figure}[ht]
\centering
  \includegraphics[width=0.49\textwidth]{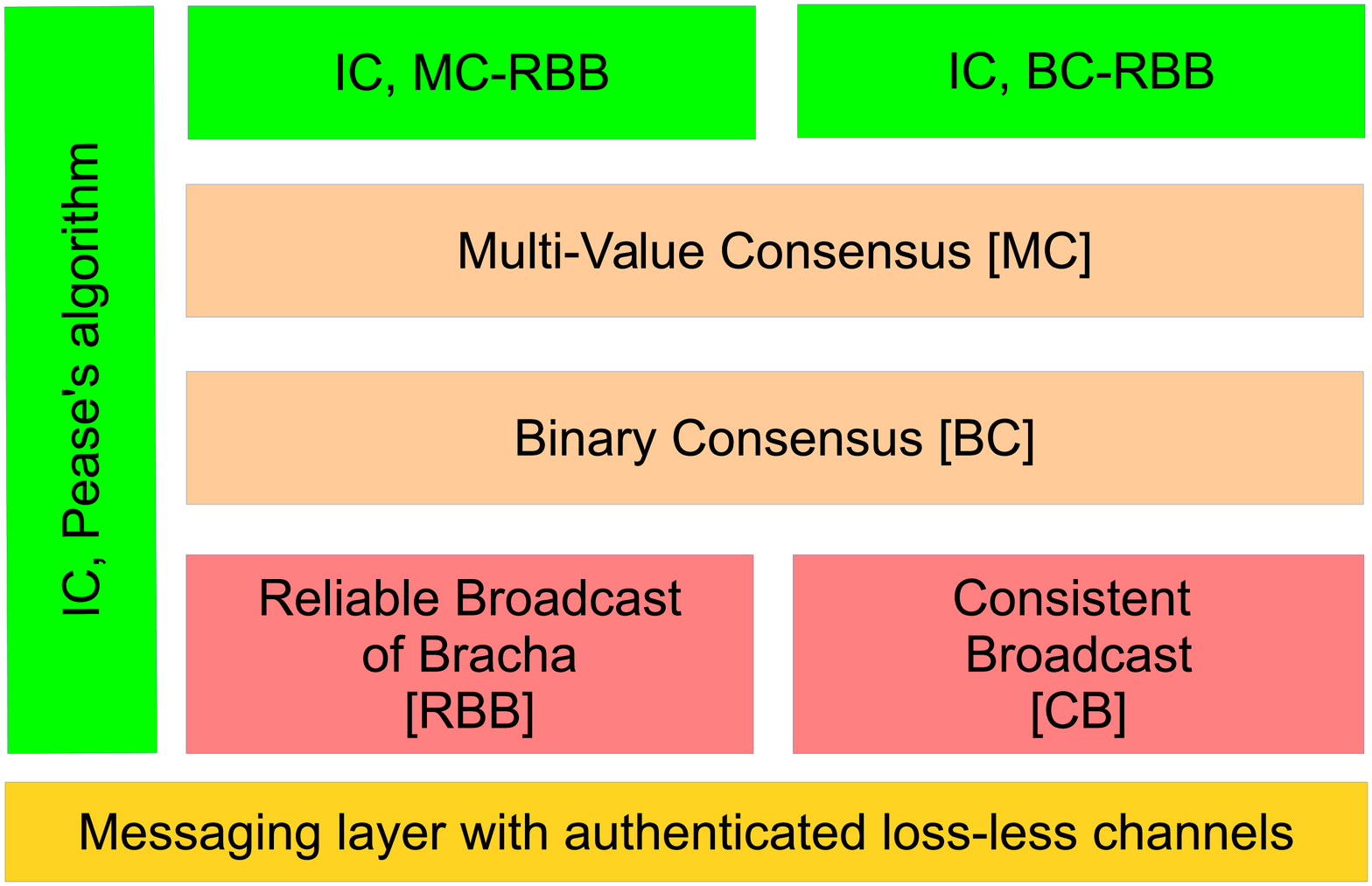}
\caption{The infrastructure and protocol stack of our implementation.}
\label{figure:stack}
\end{figure}

To reduce the overhead caused by signature operations used in Consistent Broadcast (CB), we use \emph{authenticators} as suggested by~\cite{castro}. In this scheme, nodes exchange pair-wise messages to announce to the receiving party a symmetric secret key to use when sending messages to the sending party. This exchange is repeated often enough to make the symmetric key secure. When a node wishes to multicast a message to $n$ nodes, it composes an \emph{authenticator}, which is a vector of $n$ HMACs, one for each receiving node, by using the corresponding key as input to the HMAC function. The receiving party uses its corresponding entry of the authenticator to verify both the integrity and the authenticity of the message, making this scheme analogous to digital signatures (for the closed world for which the authenticator provides HMACs). The performance improvement is vast as, in a simple evaluation on a contemporary desktop CPU, we can calculate approximately 300 SHA-1 based HMACs in the same amount of time required to produce a single digital signature using RSA with a 1024-bit key.

Bracha's binary consensus algorithm operates in a probabilistic number of phases. It requires
nodes to buffer all messages, even the ones referring to future phases to guarantee termination.
This, in conjuction with the fact that the number of rounds is not bounded, may require nodes
to buffer an arbitrary number of messages. In a practical system,
it is unrealistic to assume nodes with unbounded memory. Thus, malicious nodes could bombard
non-faulty nodes with spurious messages which, since they are required to buffer them, would
result in a state-explosion attack. Our approach on this matter is twofold. First, we identify
that each phase of Bracha's consensus protocol is independent from any previous phases. This
means that once a node enters phase $i+1$, it can safely discard any buffered messages from
phase $i$ since they are no longer needed. Second, to defend against the state-explosion
attack, nodes buffer messages whose current phase number $i$ is a total of $H$ phases ahead. However, this
requires a recovery protocol for slow nodes that have fallen behind and are unable to progress
due to the fact that the others have reached a phase $j>i+H$, which we leave as future work. 

Finally, we leverage this protocol stack to provide the following interactive consistency
algorithm suite:
\begin{enumerate}
 \item Our adaption of Pease's algorithm \emph{(IC,PEASE)}.
 \item Consistent Broadcast for the value dissemination phase and Bracha's binary consensus, in conjunction with
 the reliable broadcast of Bracha for the result consensus phase \emph{(IC,BC-RBB)}.
 \item Multicast for the value dissemination phase and multi-valued consensus, in conjunction with the reliable
 broadcast of Bracha for the result consensus phase \emph{(IC,MC-RBB)}. 
\end{enumerate}

%% file: evaluation.tex
\section{Evaluation}
\label{section:evaluation}

\begin{figure}[ht]
    \includegraphics[width=0.49\textwidth]{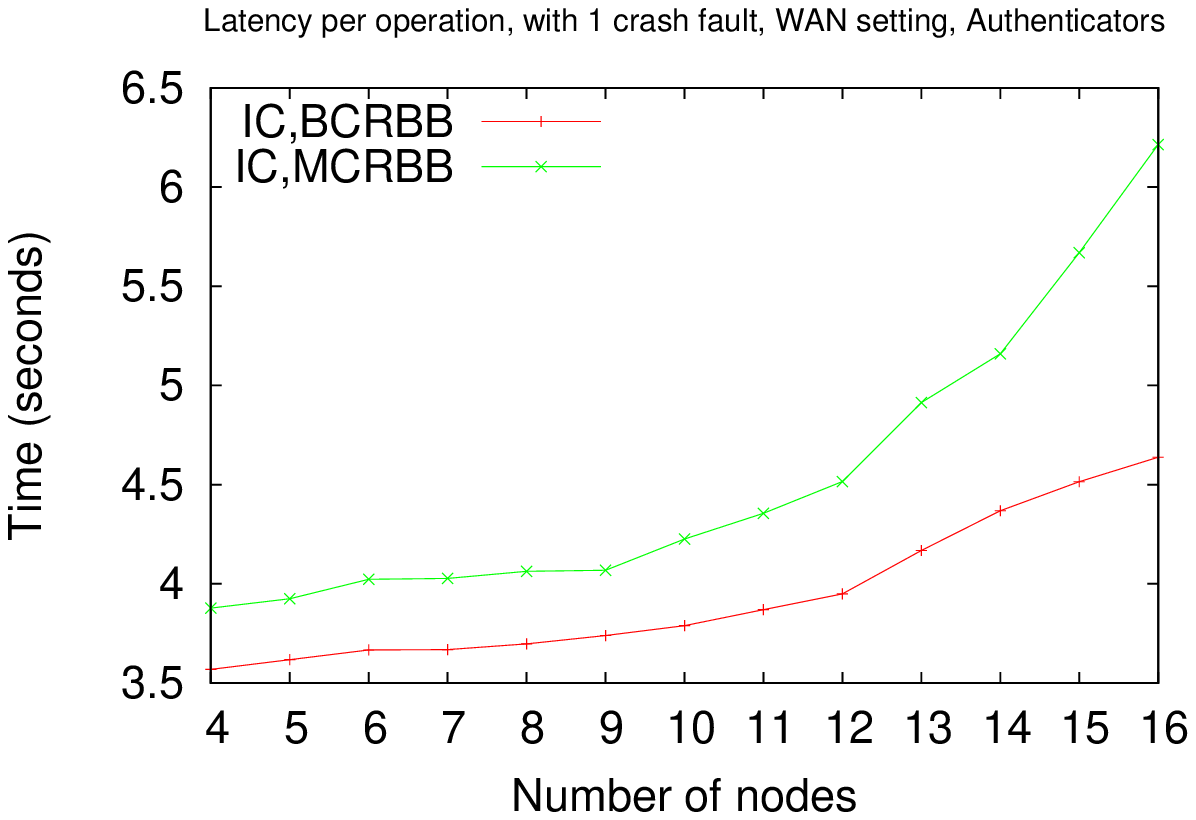}
\caption{Latency of \emph{IC,BC-RBB} and \emph{IC-MC-RBB}, with faults in WAN setting.}
\label{fig:zoominCrashWANNoLAMP}
\end{figure}

\begin{figure*}[!ht]
\centering
{
  \subfloat[]
  {
    \includegraphics[width=0.33\textwidth]{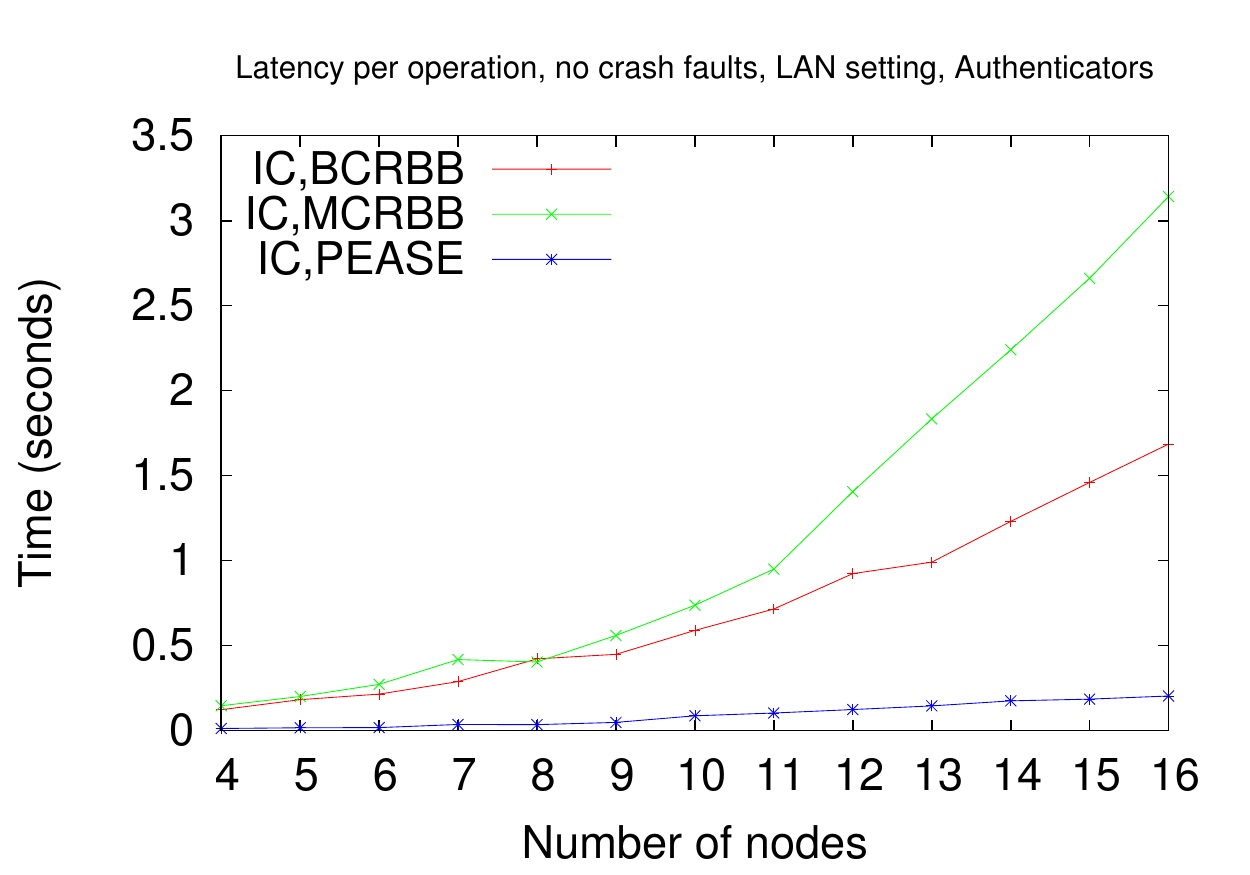}
    \label{fig:LANNoCrash}
  }
  \subfloat[]
  {
    \includegraphics[width=0.33\textwidth]{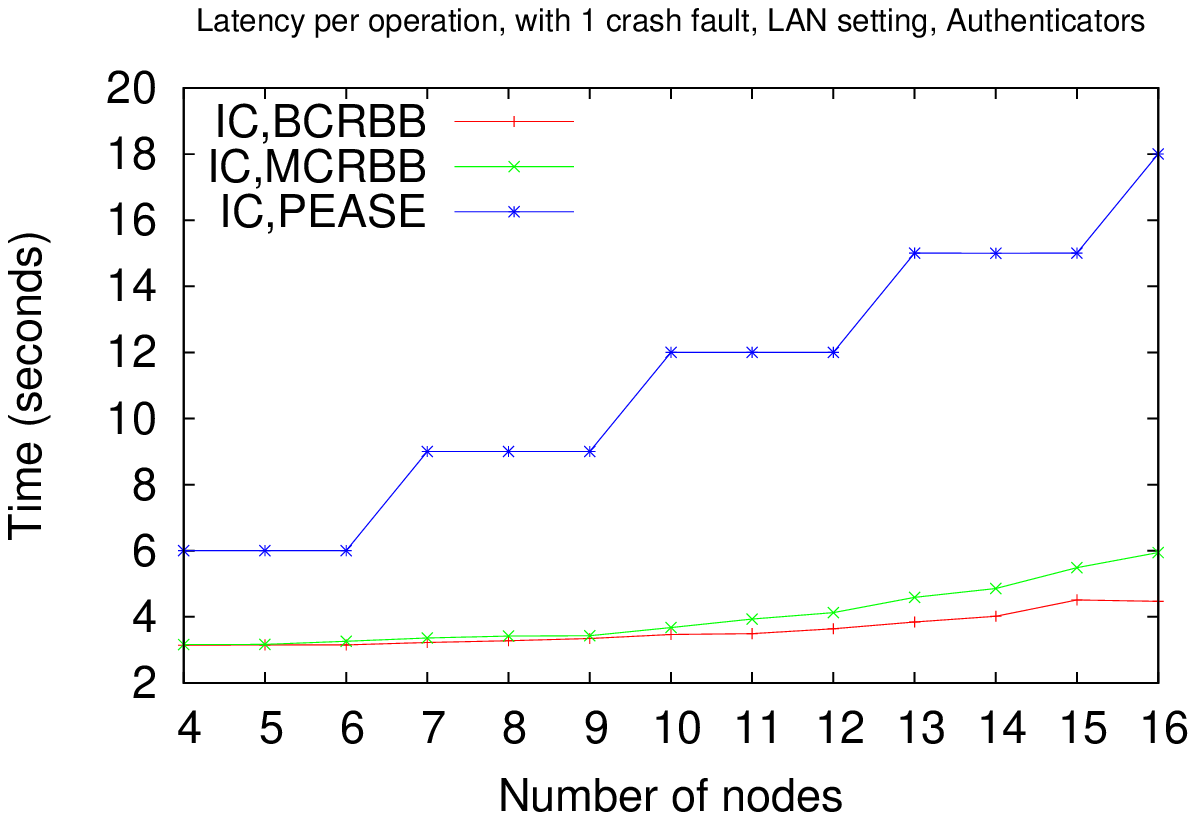}
    \label{fig:LANCrash}
  }
  \subfloat[]
  {
    \includegraphics[width=0.33\textwidth]{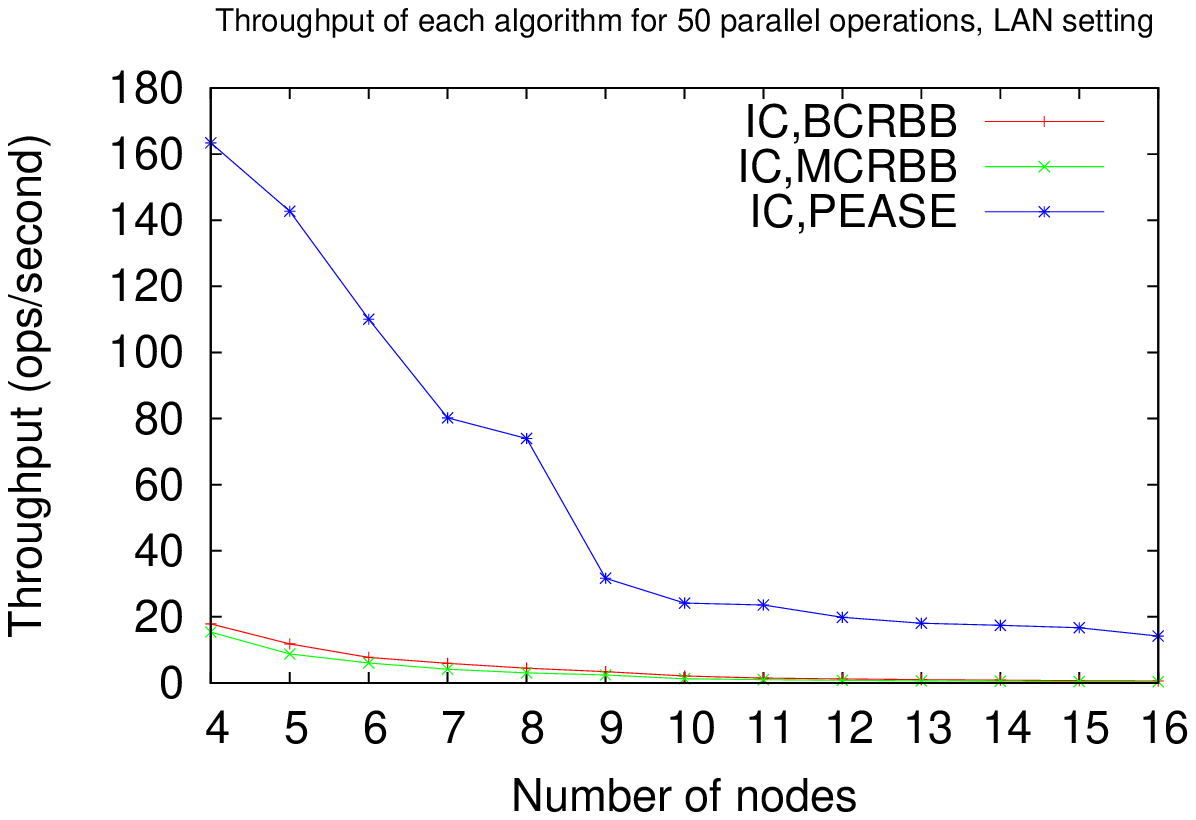}
    \label{fig:LANParallel}
  }
  \hfil
  \subfloat[]
  {
    \includegraphics[width=0.33\textwidth]{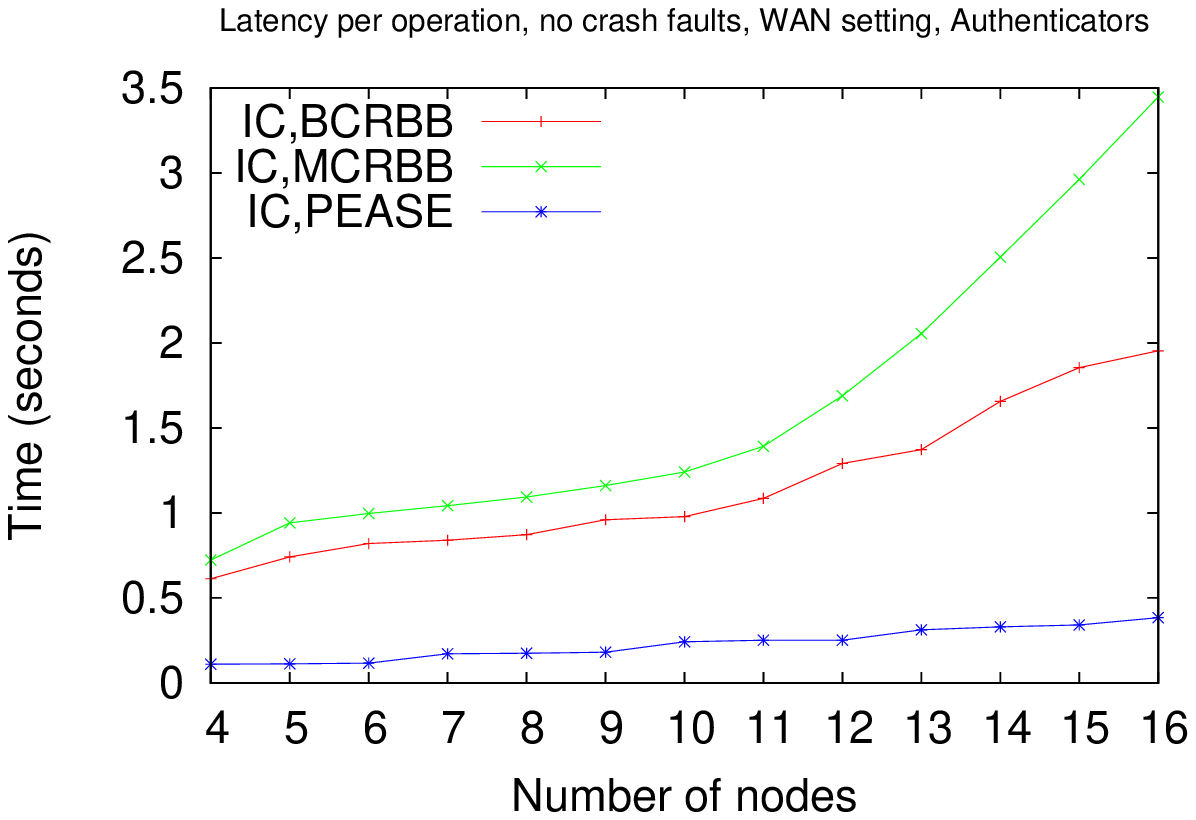}
    \label{fig:WANNoCrash}
  }
  \subfloat[]
  {
    \includegraphics[width=0.33\textwidth]{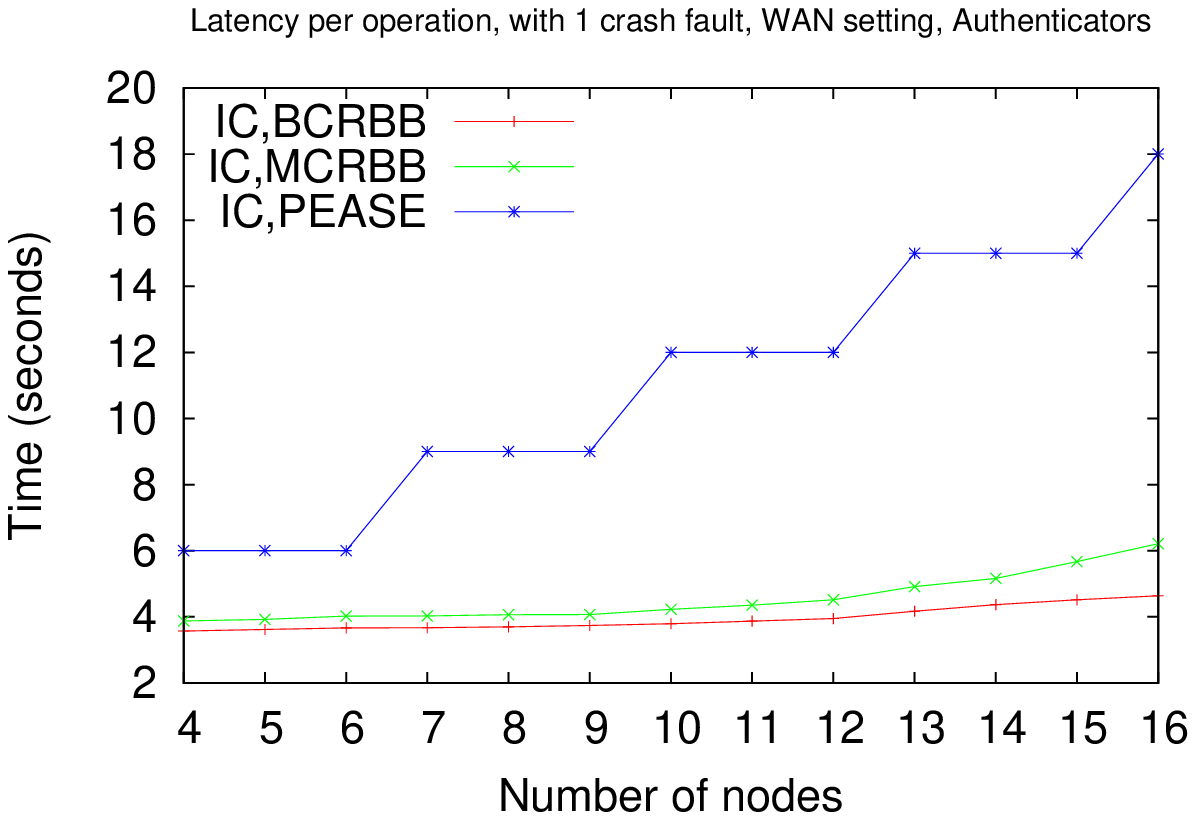}
    \label{fig:WANCrash}
  }
  \subfloat[]
  {
    \includegraphics[width=0.33\textwidth]{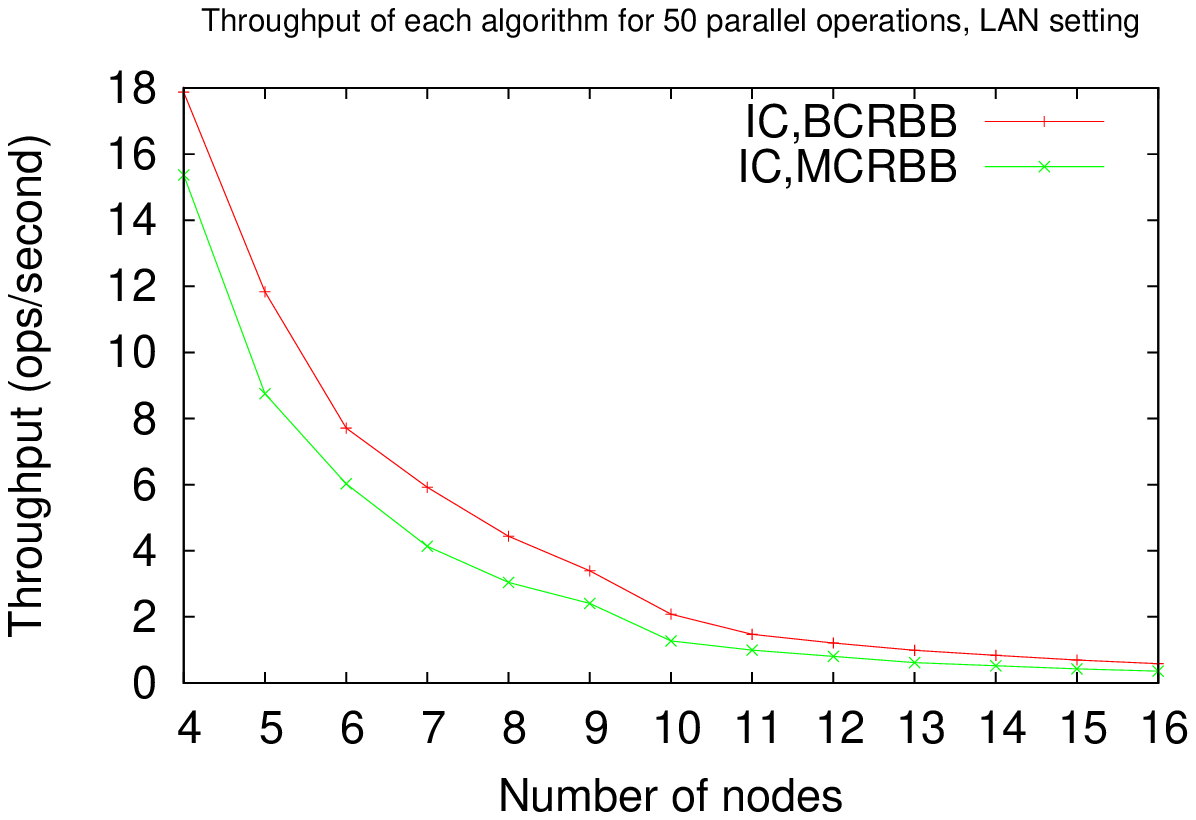}
    \label{fig:LANParallelNoLAMP}
  }
}
\caption{Performance of the algorithms under LAN (\ref{fig:LANNoCrash} - \ref{fig:LANParallel}, \ref{fig:LANParallelNoLAMP}) and WAN (\ref{fig:WANNoCrash} - \ref{fig:WANCrash})
settings. The first two columns depict the latency of each algorithm under fault-free (\ref{fig:LANNoCrash}, \ref{fig:WANNoCrash}) and faulty scenarios (\ref{fig:LANCrash}, \ref{fig:WANCrash}).
The third column depicts the throughput of each algorithm while executing 50 parallel instances with and without Pease's variant.}
\label{figure:latencyperoperationfigure}
\end{figure*}

In this section, we experimentally evaluate the performance of the 
algorithms presented above, under various system settings. We conduct our experiments using a
dedicated cluster with eight identical nodes, directly connected via an isolated 1Gbps Ethernet switch. 
Each node is configured with 4GB of RAM and dual Intel(R) Xeon(TM) CPUs
running at 2.80GHz.
We utilize up to 16 logical nodes by placing at most 2 nodes per physical machine. 
Measurements remain accurate, as cross-machine communication is always required for the algorithms to progress 
% and always dominates over communication between co-located nodes. 
% and subsumes any communication between co-located nodes.
and overshadows any communication benefits from co-located nodes. To emulate a WAN environment we utilize 
\emph{netem}~\cite{Netem}, a network emulator for Linux, to inject a uniform latency of 50ms.

In Figure~\ref{fig:LANNoCrash}, we illustrate the total time required
to complete one instance of each algorithm, without faults, in
a local area network (LAN) setting.
All algorithms are optimized to progress without waiting for timeouts, if all expected replies have been received.
Our adaptation of Pease's algorithm exhibits the best performance due to its significantly lower message complexity and the fact that no timeouts are triggered.
From the remaining alternatives, the one relying on multi-valued consensus for
the result consensus phase performs worse than its binary consensus counterpart. 
%This illustrates that the increased cost of the value dissemination phase of the IC,BC-RBB approach overweighs the two additional reliable broadcasts of the MC based approaches.
This illustrates that the \emph{(IC,BC-RBB)} approach, even though it has a more costly first phase, outperforms \emph{(IC,MC-RBB)} because of its efficiency in the second phase. We repeat this experiment in the WAN setting (Figure~\ref{fig:WANNoCrash}) and find that the same trend applies.

We now examine the effect of faults on the performance of each algorithm. We inject a single
fault in the system, which is enough to trigger timeouts and reveal the cost of timing dependencies.
In Figure~\ref{fig:LANCrash}, we illustrate the performance of each algorithm in a LAN setting, 
with a modest timeout value of three
seconds. Results illustrate the inefficiency of employing multiple timeouts in such an
environment. Our adaptation of Pease's algorithm, which is the best 
alternative in the fault-free case, actually exhibits the worst performance in the presence of a single fault. 
We repeat this experiment in a WAN environment. Our results are depicted in Figure~\ref{fig:WANCrash} 
and illustrate that the same trend applies. However, we note that we use the same timeout value in
both LAN and WAN settings, i.e., three seconds, to provide for an even comparison.
In a real deployment, one would employ a much larger timeout value for the WAN case, resulting
in a more significant impact on the algorithms' performance.
In Figure~\ref{fig:zoominCrashWANNoLAMP} we plot the same data as that shown in Figure~\ref{fig:WANCrash}, but without Pease's algorithm. This allows us to 
see more clearly the differences between the other two algorithms. We find that \emph{(IC,BC-RBB)} consistently provides latency improvements over \emph{(IC,MC-RBB)}, ranging from 10\% to 30\%.

% \begin{figure}[ht!]
%   \centering
%   \includegraphics[width=0.45\textwidth]{fault-free-LAN-latencyOneOp-ALL-AUTHENTICATORS.eps}
%   \caption{Performance benefits of authenticators over digital signatures.}
%   \label{fig:LANAuthenticators}
% \end{figure}
% 
% In all of the aforementioned cases, we used digital signatures for the algorithms that employ
% CB for the value dissemination phase and RBS as the reliable broadcast primitive of the 
% result consensus phase (IC,BC-RBB, IC,BC-RBS, IC,MV-RBS). In
% Figure~\ref{fig:LANAuthenticators}
% we illustrate the performance benefits of employing the more efficient scheme that is based
% on authenticators, as was described in Section~\ref{subsection:withsigs}. While there are no significant changes in the curves, 
% the approaches with authenticators have closed the gap with their non-signature based siblings, and even the worst performing algorithm finishes in under one second.
Lastly, we evaluate the throughput of the algorithms by running 50 parallel instances
of each algorithm, in the LAN setting. We present our results in
Figure~\ref{fig:LANParallel} (all algorithms) and~\ref{fig:LANParallelNoLAMP} (without Pease's variant).
Since it is a fault-free case, Pease's variant exhibits the
best performance. Between \emph{(IC,BC-RBB)} and \emph{(IC,MC-RBB)}, the former consistently provides higher throughput ranging from 16\% to 63\% improvement. 
\emph{(IC,BC-RBB)}'s improved performance results from the reduced complexity of the result consensus phase, which is achieved by the use of consistent broadcast in the value dissemination phase. We repeated the same experiments in the WAN setting and observe similar trends, which we omit due to lack of space.
% For the remaining alternatives, results illustrate that \emph{(IC,BC-RBB)} that
% employs CB for the value dissemination phase, thus requiring only one BC instance for the
% result consensus phase, provides consistently better throughput. 
% Thus, by strengthening the value dissemination phase, we improve performance.

%To summarize, in Table~\ref{table:ranking}, we present the ranking of the implemented algorithms across different runtime scenarios.
To summarize, our adaptation of Pease's variant is fastest in fault-free settings, due to its reduced message complexity and its lack of signature operations. However, once faults are introduced, \emph{(IC,BC-RBB)} performs the best. Furthermore, \emph{(IC,BC-RBB)} is among the top-two for all evaluated scenarios and requires only one synchronization point, in contrast with Pease's variant, which requires multiple. The reader will notice that absolute performance numbers are low. However, we are not proposing IC as a means to implement high-throughput applications. Instead, we have found IC to be useful in applications, such as a distributed e-voting system we built. We use it there, only once per election, to resolve diverged views of multiple vote collectors, after the election has completed~\cite{ddemos}. 
%after voting is finished, while, during voting, a much lighter protocol is used.

% \begin{table}[!ht]
%   \renewcommand{\arraystretch}{1.3}
%   \centering
%   \begin{tabular}{c|c|c}
%     & \bfseries \emph{t} = 0 & \bfseries \emph{t} $>$ 0\\
%     \hline
%     \bfseries LAN & \begin{tabular}{@{}c@{}} \emph{IC, LAMP}\\ \emph{IC, BC-RBB}\\ \emph{IC, MC-RBB}\\ \emph{IC, BC-RBS}\\ \emph{IC, MC-RBS}\end{tabular} & \begin{tabular}{@{}c@{}} \emph{IC, BC-RBB}\\ \emph{IC, MC-RBB}\\ \emph{IC, BC-RBS}\\ \emph{IC, MC-RBS}\\ \emph{IC, LAMP}\end{tabular}\\
%     \hline
%     \bfseries WAN & \begin{tabular}{@{}c@{}} \emph{IC, LAMP}\\ \emph{IC, BC-RBS}\\ \emph{IC, BC-RBB}\\ \emph{IC, MC-RBS}\\ \emph{IC, MC-RBB}\end{tabular} & \begin{tabular}{@{}c@{}} \emph{IC, LAMP}\\ \emph{IC, BC-RBS}\\ \emph{IC, BC-RBB}\\ \emph{IC, MC-RBS}\\ \emph{IC, MC-RBB}\end{tabular}\\
%     \hline
%   \end{tabular}
%   \vspace{10px}
%   \caption{Ranking of algorithms for different runtime scenarios.}
%   \label{table:ranking}
% \end{table}

%% file: conclusion.tex
\section{Conclusion}
\label{section:conclusion}
In this paper, we tackle the problem of Interactive Consistency (IC) in practical, real-world systems. This problem has received little attention in this setting so far, and we present a suite of algorithms and their implementations which can be used to solve this problem. These range from porting Pease's synchronous algorithm and making multiple timing assumptions, to composing more sophisticated algorithms based on existing broadcast and consensus primitives with a single timing assumption.
%We provide a signature-based reliable broadcast primitive, which can substitute existing signature-free incarnations of reliable broadcast.
We also define a more relaxed version of the problem, which we call \emph{Eventual Interactive Consistency} (EIC), that is suitable for some applications, and we describe possible approaches for solving the problem without any timing assumptions.

Most of the algorithms in the protocol stack we built have been proposed but never been implemented and evaluated (to our knowledge) in a real system before. We have experimentally compared the performance of all algorithms and highlighted trade-offs that arise in different system settings. We find that one size does not fit all; for example, our adaptation of Pease's algorithm appears to be more appropriate for settings where node failures are rare, but once faults do occur, its performance degrades more than the remaining approaches. 
%Thus, system designers need to carefully consider network factors and likelihood of faults before choosing which IC approach to apply. 
With this work, we hope to provide a framework with which system designers can reason about the appropriate IC approach to use. A download link for our open-source software will be included upon publication.

%% file: icproof.tex
%\onecolumn
% \documentclass[conference, 10pt]{IEEEtran}
% \usepackage[cmex10]{amsmath}
% \usepackage{amsfonts}
% \usepackage{algorithmic}
% \usepackage[caption=false,font=footnotesize]{subfig}
% \usepackage{fixltx2e}
% \usepackage{multirow}
% \usepackage[table]{xcolor}
% 
% \begin{document}

% \appendices

\section{Proof of Correctness - Interactive Consistency}
\label{section:appendix1proof}
\newtheorem{theorem}{Theorem}
\newtheorem{lemma}{Lemma}
\newtheorem{corollary}{Corollary}[theorem]

\newenvironment{myproof}[1][Proof:]{\begin{trivlist}
\item[\hskip \labelsep {\bfseries #1}]}{\end{trivlist}}

\subsection{The adversarial model}
\label{subsection:adversary}

We assume a distributed system of $n$ nodes that communicate with each other via authenticated message exchanges.
Each node $n_i$ has an arbitrary length private value $v_i \in \mathfrak{U}$, where $\mathfrak{U}$ is the domain of values that can be proposed.
We assume that there exists a global clock variable $Clock \in \mathbb{N}$ and that each node
$n_i$ is equipped with a local clock variable $Clock[n_i] \in \mathbb{N}$ and an input message tape. When node $n_i$ wishes to send
a message $m$ to node $n_j$, it sends a tuple $(n_i,n_j,m)$ to the adversary $A$.

To model synchronization loss among the clocks of honest nodes caused by the adversary $A$, we define the following events:
\begin{enumerate}
\item The event $Inc(i): i \leftarrow i+1$ that causes clock $i$ to advance by one time unit.
\item The event $Init(n_i): Clock[n_i] \leftarrow Clock$ that initializes the local clock variable of a node $n_i$ by synchronizing it
with the global clock variable $Clock$.
\end{enumerate}

% Nodes are assumed to execute (in order) the following abstract procedures: $Initialize()$, $Value\_Dissemination()$ and
% $Result\_Consensus()$. 
% 
% In Figure~\ref{figure:initanddisseminate}, we illustrate (in order) the procedures that the nodes employ for the first
% phase of our IC protocols. The $Value\_Dissemination()$ procedure encompasses a $Disseminate()$ primitive. This primitive
% provides an abstraction of the algorithms that the nodes use to disseminate their private value to the rest of the nodes in
% the system. Depending on the IC algorithm at hand, this can either refer to a simple multicast, or a consistent
% broadcast (CB). The $Disseminate()$ primitive receives as input two parameters, the unique identifier of node $n_i$ and its private
% value $v_i$. Several instances of the $Disseminate()$ primitive may be active at the same time. The inclusion of the node's unique identifier
% allows us to distinguish between messages sent in different instances, as well as forming tuples $(n_i,n_j,m)$ for
% handing to the adversary $A$. When an instance of a $Disseminate()$ primitive corresponding to $n_j$ concludes to some node $n_i$, we assume
% that a $(n_j,v_j)$ tuple is $delivered$ to $n_i$.

The adversarial setting for $A$ is as follows:
\begin{enumerate}
 \item The adversary $A$ can corrupt a fixed subset of nodes $M$, where $max\{|M|\}=t \leq \lfloor \frac{n-1}{3} \rfloor$. 
 If $A$ adds a node $n_i$ to $M$, then it has full control over it. If $n_i \notin M$, then $A$ runs once, and in order, the
 procedures $Init(n_i)$ and $Initialize()$ to successfully initialize $n_i$. Then, $n_i$ will spawn a task T1 which will execute the $Value\_Dissemination()$ procedure.
 \item Access to the local clock variable of $n_i$ is restricted to $A$ and $n_i$. No honest node
 has access to the global clock variable $Clock$.
 \item $A$ may arbitrarily invoke the events $Inc(Clock[n_i])$ and $Inc(Clock)$.
 \item There exists an upper bound $\Delta$ on the drift of all honest nodes' local clocks compared to the
 global clock variable, i.e., $|Clock[n_i] - Clock| \leq \Delta$, where $|\cdot|$ denotes the absolute value.
 \item There exists a value $End$ such that for each honest node $n_i$, if $Clock[n_i] \geq End$, then, $n_i$ assumes that
 the value dissemination phase has ended and will immediately move on to the result consensus phase, i.e., it will spawn a
 task T2 that will immediately start executing the $Result\_Consensus()$ procedure. At this point, task T1
 will immediately drop, and will not process, messages of the value dissemination phase.
 \item $A$ can freely examine the contents of any message $m$ of any tuple $(n_i,n_j,m)$.
 \item $A$ can write on the input message tape of any node $n_i$.
 \item \label{item:advmessagedelivery} There exists an upper bound $\delta$ on the time that $A$ can delay the delivery of the messages between
 honest nodes. More formally, when an honest node $n_i$ sends a tuple $(n_i,n_j,m)$ to $A$, if the value of the
 global clock variable is $t$, then $A$ must write $(n_i,m)$ to the input tape of honest node $n_j$ by global clock time:
 $Clock=t+\delta$.
\end{enumerate}

For the first phase of our IC protocols, we will derive an upper bound on the time
required for honest nodes to receive the private values of all other honest nodes. This bound will be derived using the upper time
bounds that correspond to each step of the value dissemination phase. The upper time bound of each step will be computed based on
the appropriate reference point. For instance, when we compute an upper time bound
of a node's local computation, the reference point will be a node's local clock variable $Clock[n_i]$. Similarly, when we compute an upper time
bound on the delivery of a message between honest nodes, the reference point will be the global clock variable $Clock$, according to
assumption~\ref{item:advmessagedelivery} above.

% The complete algorithms of all the abstract procedures that honest nodes execute during our IC protocols are
% given in their corresponding section.

\subsection{IC,MC-RBB - Proof of Correctness}
\label{subsection:icmcrbbproof}

\begin{figure}[ht!]
  \footnotesize
  \begin{algorithmic}
  \STATE \textbf{Common node inputs:} $n$, $t \leq \lfloor \frac{n-1}{3} \rfloor$, $End$
  \STATE \textbf{Local inputs for node $n_i$:} $n_i$, $v_i$
  \STATE \textbf{Local output for node $n_i$:} $V$
  \\
  \null
  \STATE \textbf{procedure} $Initialize():$
    \FOR{$1 \leq i \leq n$}
      \STATE $V[i] = \perp$;
    \ENDFOR
    \STATE $agreedValues = 0$;
  \\
  \null
  Task T1:
  \STATE \textbf{procedure} $Value\_Dissemination():$
  \STATE $multicast(n_i,v_i)$;
  \WHILE{1}
    \STATE wait until $(n_j,v_j)$ is $obtained$;
    \STATE $V[j] = v_j$;
  \ENDWHILE
  \\
  \null
  Task T2:
  \STATE \textbf{procedure} $Result\_Consensus():$
    \FOR{$1 \leq i \leq n$}
      \STATE $MV\_Consensus(i,V[i])$
    \ENDFOR
    \WHILE{$agreedValues \neq n$}
      \STATE wait until $(n_i,v_i)$ is $agreed$;
      \STATE $V[i]=v_i$;
      \STATE $++agreedValues$;
    \ENDWHILE
  \end{algorithmic}
  \caption{Protocol \emph{(IC,MC-RBB)}.}
  \label{figure:icmcrbbabstractalgorithms}
\end{figure}

In this section we prove the correctness of our \emph{(IC,MC-RBB)} algorithm. Figure~\ref{figure:icmcrbbabstractalgorithms}
presents the algorithm of this protocol. Recall that in \emph{(IC,MC-RBB)}, nodes use a simple
multicast for the value dissemination phase and $n$ parallel instances of multi-valued consensus for the result consensus
phase. A $multicast()$ instance receives two inputs. The first input is the unique identifier of the node initiating the $multicast()$. The second
input is the value that will be multicast. When an instance of a $multicast()$ primitive initiated by $n_j$ terminates at
some node $n_i$, we state that an $(n_j,v_j)$ tuple is $obtained$ by $n_i$. The $Result\_Consensus()$ procedure invokes
an $MV\_Consensus()$ primitive. This is a multi-valued consensus protocol and it receives two inputs. The first input is a consensus instance
identifier $i$, where $i \in [1,n]$. The second input is the value that will be used as the private input for this consensus instance
by the node invoking $MV\_Consensus()$. When $MV\_Consensus()$ instance $i$ terminates at some node $n_j$, we say that an $(n_i,v_i)$ tuple is $agreed$
and we assume that this tuple becomes available to $n_j$.

\begin{table}[ht!]
\centering
\begin{tabular}{ | >{\centering\arraybackslash}m{2.5cm} | >{\centering\arraybackslash}m{2.5cm} | >{\centering\arraybackslash}m{2.5cm} |}
  \hline
  \multirow{2}{*}{Algorithmic Step} & \multicolumn{2}{c|}{Upper time bound of each clock} \\
  \cline{2-3}
  & $Clock$ & $Clock[n_i]$ \\ 
  \hline
  Honest nodes are initialized & \cellcolor{gray} 0 & $\Delta$ \\
  \hline
  Honest nodes multicast their values & $t_{comp} + 2\Delta$ & \cellcolor{gray} $t_{comp} + \Delta$ \\
  \hline
  Honest nodes receive the private values of honest nodes & \cellcolor{gray} $t_{comp} + 2\Delta + \delta$ & $t_{comp} + 3\Delta + \delta$ \\
  \hline
\end{tabular}
\caption{Upper time bounds for $Clock$ and $Clock[n_i]$ at each algorithmic step. Grayed cells indicate the reference point (clock variable)
with respect to which upper time bounds are calculated.}
\label{table:icmcrbbclocks}
\end{table}

\begin{theorem}
Let $A$ be an adversary model as the one described in Appendix~\ref{subsection:adversary} and that all honest nodes have been
successfully initialized at global clock time: $Clock = 0$. Let $t_{comp}$ be the maximum preparation time required to
multicast a message. Every honest node $n_i$ will have $obtained$ all of the private values of all honest nodes
when its local clock reading is: $Clock[n_i] \leq t_{comp} + 3\Delta + \delta$.
\label{theorem:phase1icmcrbb}
\end{theorem}
\begin{myproof}
We will compute an upper bound on the time required for honest nodes to receive the private values of all other honest nodes.
Table~\ref{table:icmcrbbclocks} illustrates the upper time bounds of all clock variables, at each algorithmic step, to ease
the understanding of the computation described below.

The initialization of every honest node $n_i$ will have been completed at local clock time: $Clock[n_i] \leq \Delta$.
Then, each honest node $n_i$ performs at most $t_{comp}$ steps before it multicasts its value to the rest of the nodes.
When each honest node $n_i$ multicasts its value to the rest of the nodes, the value of the global clock is: $Clock \leq t_{comp} + 2\Delta$.
All honest nodes will receive the private values of all other honest nodes at global clock time: $Clock \leq t_{comp} + 2\Delta + \delta$, which implies that 
the time of their internal clocks is at most $t_{comp} + 3\Delta + \delta$.
\end{myproof}

\begin{corollary}
 By setting $End \geq t_{comp} + 3\Delta + \delta$, if $n_i$ is non-faulty, then
 all non-faulty nodes will have $obtained$ its private value $v_i$, by the beginning of the result consensus phase.
\label{corollary:icmcrbb1}
\end{corollary}

We will use Corollary~\ref{corollary:icmcrbb1} to prove
two Lemmas that will be used, subsequently, to prove that \emph{(IC,MC-RBB)} satisfies all of the IC properties.

\begin{lemma}
If at the beginning of the result consensus phase of \emph{(IC,MC-RBB)} all non-faulty nodes
know the same value $v_i$ for node $n_i$, then $v_i$ will appear in the $i^{th}$ slot
of the result vector $V$ of all non-faulty nodes.
\label{lemma:phase21icmcrbb}
\end{lemma}
\begin{myproof}
 From the \emph{MVC1 Validity}
 property of multi-valued consensus, we have that every correct process that decides, will
 decide $v_i$. Due to the \emph{MVC5 Termination} property of multi-valued consensus, all non-faulty
 nodes eventually decide. Thus, all correct processes will eventually decide $v_i$ and will set
 $V[i]=v_i$.
\end{myproof}

\begin{lemma}
If at the beginning of the result consensus phase of \emph{(IC,MC-RBB)} all non-faulty nodes
have conflicting opinions about the value of (faulty) node $n_i$, then some value $v_i$ will appear in the
$i^{th}$ slot of the result vector $V$ of all non-faulty nodes.
\label{lemma:phase22icmcrbb}
\end{lemma}
\begin{myproof}
 Due to the \emph{MVC5 Termination} property of multi-valued consensus, all non-faulty nodes will eventually decide on a value.
 From the \emph{MVC4 Agreement} property of multi-valued consensus, we have that correct processes never decide differently.
 Thus, all correct processes will decide on some $v_i$ and will set $V[i]=v_i$.
\end{myproof}

% Now, we will use Lemmas \ref{lemma:phase21} and \ref{lemma:phase22} to prove that all of the IC properties are
% satisfied by our algorithms.

\begin{theorem}
 The \emph{IC,MC-RBB} protocol satisfies the properties of IC.
\end{theorem}
\begin{myproof}
 
 \emph{Agreement}: Lemmas \ref{lemma:phase21icmcrbb} and \ref{lemma:phase22icmcrbb} guarantee that all non-faulty nodes will always
 agree on the same value $v_i$ for the $i^{th}$ slot of the result vector, $\forall i \in [1,n]$.
 Thus, since every slot of the result vector will have the same value across all non-faulty nodes,
 it follows that all non-faulty nodes agree on the same vector of values $V$.
 
 \emph{Validity}: This follows directly from Lemma~\ref{lemma:phase21icmcrbb}.
 
%  This property
%  discriminates between two cases, i.e., whether $n_i$ is faulty or not.
%  Assume that $n_i$ is faulty. We have to prove that all non-faulty nodes will agree on any value $v_i$ for
%  $V[i]$. This follows directly from Lemma~\ref{lemma:phase22icmcrbb}.
%  Assume that $n_i$ is non-faulty. We have to prove that all non-faulty nodes will agree on the same value
%  $V[i]=v_i$, where $v_i$ is the private value of $n_i$. 
 
 \emph{Termination}: From Lemmas \ref{lemma:phase21icmcrbb} and \ref{lemma:phase22icmcrbb}, it follows that all non-faulty
 nodes are guaranteed to agree on a value for each slot of the result vector $V$. Thus, all non-faulty nodes
 will eventually decide on a vector $V$.
\end{myproof}

\subsection{Consistent broadcast}
\label{subsection:cbdescription}

\begin{figure}[ht!]
   \centering
   \footnotesize
   \begin{algorithmic}
    \STATE \textbf{Initialization:}
      \begin{itemize}
       \item[] $\tilde{v} \gets \bot$; $\tilde{\nu} \gets \bot$
       \item[] $v‎' \gets \bot$
       \item[] $W \gets \emptyset$; $r \gets 0$
      \end{itemize}
    \STATE
    \STATE Upon \textit{c-broadcast($N_i$, $v$)}:
      \begin{itemize}
	\item[] $v‎' \gets v$
	\item[] send message ($c\textrm{-}send$, $N_i$, $v$) to all
      \end{itemize}
    \STATE  
    \STATE Upon receiving message ($c\textrm{-}send$, $N_b$, $v$) from $N_s$:
       \begin{itemize}
	  \item[] \textbf{if} $s = b$ \textbf{and} $\tilde{v} = \bot$ \textbf{then}
	  \begin{itemize}
	    \item[] $\tilde{v} \gets v$
	    \item[] compute signature $\sigma$ on ($c\textrm{-}ready$, $N_b$, $v$)
	    \item[] send message ($c\textrm{-}ready$, $N_b$, $v$, $\sigma$) to $N_s$
	  \end{itemize}
       \end{itemize}
    \STATE
    \STATE Upon receiving message ($c\textrm{-}ready$, $N_b$, $v$, $\sigma$) from $N_s$ for the first time:
      \begin{itemize}
	\item[] \textbf{if} $b = i$ \textbf{and} $v = v'$ \textbf{and} $\sigma$ is a valid signature \textbf{then}
	\begin{itemize}
	  \item[] $W \gets W \cup \{\sigma\}$
	  \item[] $r \gets r + 1$
	  \item[] \textbf{if} $ r = n - t$ \textbf{then}
	  \begin{itemize}
	    \item[] send message ($c\textrm{-}final$, $N_i$, $v$, $W$) to all
	  \end{itemize}
	\end{itemize}
      \end{itemize}
    \STATE
    \STATE Upon receiving message ($c\textrm{-}final$, $N_b$, $v$, $W$) from $N_s$:
      \begin{itemize}
       \item[] \textbf{if} $\tilde{\nu} = \bot$ \textbf{and} $W$ contains $n - t$ valid signatures for $v$ \textbf{then}
       \begin{itemize}
	  \item[] $\tilde{\nu} \gets v$
	  \item[] \textit{c-deliver($v$)}
	\end{itemize}
      \end{itemize}
  \end{algorithmic}
  \caption{Consistent Broadcast using digital signatures for node $N_i$.}
  \label{fig:cb}
\end{figure}

In Figure~\ref{fig:cb}, we present a simplified adaption of Consistent Broadcast from~\cite{cachin2001secure}. In this protocol,
each source node $n_i$ sends its value $v$ to all other nodes and waits for $n - t$ signed endorsement responses. Each recipient node
endorses only the first value received for every broadcast. Once the corresponding responses are accumulated, the sender encloses
all received endorsements in a uniqueness certificate and sends the certificate, along with the value, to all nodes of the system.
Consistent Broadcast delivers a value $v$, \emph{iff} the associated certificate contains at least $n - t$ valid signatures.

\subsection{IC,BC-RBB - Proof of Correctness}
\label{subsection:icbcrbbproof}

In this section we prove the correctness of our \emph{(IC,BC-RBB)} algorithm. Figure~\ref{figure:icbcrbbabstractalgorithms} presents
the algorithm of this protocol. Recall that in \emph{(IC,BC-RBB)}, nodes use
consistent broadcast for the value dissemination phase and $n$ parallel instances of binary consensus for the result
consensus phase. In contrast to the original implementation of CB that accounts for membership changes across different system views, our IC protocols
assume static membership, i.e., there is only one view. Thus, we can safely strip the view number from the messages of CB.
We denote
a message $m$ signed by node $n_i$ as ${\langle m \rangle}_{\sigma_i}$. A $Consistent\_Broadcast()$
instance receives as inputs two values. The first input is the unique identifier of the node initiating the $Consistent\_Broadcast()$ instance. The second
input is the value that will be broadcast. When an instance of $Consistent\_Broadcast()$ initiated by $n_j$ terminates to
some node $n_i$, it $delivers$ an $(n_j,v_j,c_j)$ tuple, which we say is $obtained$ by $n_i$. The third value, $c_j$, of an $obtained$ tuple $(n_j,v_j,c_j)$
is a set of $\langle n_j,v_j \rangle_{\sigma_k}$ tuples with valid signatures, from $(n-t)$ distinct nodes that support the claim
that $v_i$ is the private value of $n_i$. The $Result\_Consensus()$ procedure invokes
a $B\_Consensus()$ primitive. This is a binary consensus protocol and it receives two inputs. The first input is a consensus instance
identifier $i$, where $i \in [1,n]$. The second input is a binary value that will be used as the private input for this consensus instance
by the node invoking $B\_Consensus()$. When $B\_Consensus()$ instance $i$ terminates at some node $n_j$, we say that an $(n_i,b_i)$ tuple is $agreed$
and we assume that this tuple becomes available to $n_j$. We say that the outcome of a binary consensus instance $i$ is the value $b_i$ of an
$agreed$ tuple $(n_i,b_i)$.

\begin{theorem}
Let $A$ be an adversary model as the one described in Appendix~\ref{subsection:adversary} and that all honest nodes have been
successfully initialized at global clock time: $Clock = 0$. Let $t_{comp}$ be the worst-case running time, across all nodes, of
any local computation required at any step of CB, including the maximum time required to prepare to multicast and/or send a message.
Every honest node $n_i$ will have $obtained$ all of the private values of all honest nodes
when its local clock reading is: $Clock[n_i] \leq (n+2)t_{comp} + 7\Delta + 3\delta$.
\label{theorem:phase1icbcrbb}
\end{theorem}
\begin{myproof}
We will compute an upper bound on the time required for honest nodes to receive the private values of all other honest nodes.
Table~\ref{table:icbcrbbclocks} illustrates the upper time bounds of all clock variables, at each algorithmic step, to ease
the understanding of the computation described below.

The initialization of every honest node $n_i$ will have been completed at local clock time: $Clock[n_i] \leq \Delta$.
Then, each honest node $n_i$ performs at most $t_{comp}$ steps before it multicasts an \emph{c-send} message to the rest of the nodes.
When each honest node $n_i$ multicasts an \emph{c-send} message to the rest of the nodes, the value of the global clock is: $Clock \leq t_{comp} + 2\Delta$.
All honest nodes will receive the \emph{c-send} messages of all other honest nodes at global clock time: $Clock \leq t_{comp} + 2\Delta + \delta$, which
implies that their local clock variable is at most: $Clock[n_i] \leq t_{comp} + 3\Delta + \delta$. Apparently, each honest node will receive,
at most, $n$ \emph{c-send} messages. Therefore, each honest node $n_i$ will perform at most $n \times t_{comp}$ steps before it responds with an \emph{c-ready}
message for each \emph{c-send} message it received. Each honest node will receive the \emph{c-ready} messages for its own private value at global
clock time: $Clock \leq (n+1)t_{comp} + 4\Delta + 2\delta$. Then, correct nodes will prepare to multicast a \emph{c-final} message for their
private value, which will require, at most, $t_{comp}$ steps. This will take place when their local clock reading is:
$Clock[n_i] \leq (n+2)t_{comp} + 5\Delta + 2\delta$. All honest nodes will receive the \emph{c-final} messages for the private values
of every other honest node at global clock time: $Clock \leq (n+2)t_{comp} + 6\Delta + 3\delta$, which implies that their local clock
reading is: $Clock[n_i] \leq (n+2)t_{comp} + 7\Delta + 3\delta$. Thus, every honest node will $obtain$ the private value of every other
honest node by local clock time: $Clock[n_i] \leq (n+2)t_{comp} + 7\Delta + 3\delta$.
\end{myproof}

\begin{figure}[ht!]
  \footnotesize
  \begin{algorithmic}
  \STATE \textbf{Common node inputs:} $n$, $t \leq \lfloor \frac{n-1}{3} \rfloor$, $End$
  \STATE \textbf{Local inputs for node $n_i$:} $n_i$, $v_i$
  \STATE \textbf{Local output for node $n_i$:} $V$
  \\
  \null
  \STATE \textbf{procedure} $Initialize():$
    \FOR{$1 \leq i \leq n$}
      \STATE $V[i] = \perp$;
      \STATE $C[i] = \perp$;
    \ENDFOR
    \STATE $agreedValues = 0$;
  \\
  \null
  Task T1:
  \STATE \textbf{procedure} $Value\_Dissemination():$
  \STATE $Consistent\_Broadcast(n_i,v_i)$;
  \WHILE{1}
    \STATE wait until $(n_j,v_j,c_j)$ is $obtained$;
    \STATE $V[j] = v_j$;
    \STATE $C[j] = c_j$;
  \ENDWHILE
  \\
  \null
  Task T2:
  \STATE \textbf{procedure} $Result\_Consensus():$
    \FOR{$1 \leq i \leq n$}
      \IF{$V[i] = \perp$}
        \STATE $B\_Consensus(i,0)$;
      \ELSE
        \STATE $B\_Consensus(i,1)$;
      \ENDIF
    \ENDFOR
    \WHILE{ $agreedValues \neq n$}
      \STATE wait until $(n_i,b_i)$ is $agreed$;
      \STATE $\mathfrak{B}[i] = b_i$;
      \STATE $++agreedValues$;
    \ENDWHILE
    \STATE $Finalize\_V()$;
  \end{algorithmic}
  \caption{Protocol \emph{(IC,BC-RBB)}.}
  \label{figure:icbcrbbabstractalgorithms}
\end{figure}

\begin{corollary}
 By setting $End \geq (n+2)t_{comp} + 7\Delta + 3\delta$, if $n_i$ is non-faulty, then
 all non-faulty nodes will have $obtained$ its private value $v_i$, by the beginning of the result consensus phase.
\label{corollary:icbcrbb1}
\end{corollary}

% From Theorem~\ref{theorem:phase1icbcrbb} it immediately follows that by setting $End \geq (n+2)t_{comp} + 7\Delta + 3\delta$, at the 
% beginning of the result consensus phase, if $n_i$ is non-faulty, then all non-faulty nodes have $delivered$ its private value $v_i$.
In the result consensus phase of \emph{(IC,BC-RBB)}, an honest node $n_i$ will vote 1 for binary consensus instance $j$ \emph{iff} it
$obtained$ the private value of $n_j$ by the end of the value dissemination phase. Otherwise, an honest node will always vote 0.
We assume that all non-faulty nodes store the outcome of every binary consensus instance $i$ in the $i^{th}$ slot of a
vector $\mathfrak{B}$, where $\mathfrak{B}[i] \in \{0,1\}$, $\forall i \in [1,n]$. Vector $\mathfrak{B}$ should not be confused
with vector $V$, which is the result vector.

\begin{lemma}
 All non-faulty nodes will eventually agree on the same vector $\mathfrak{B}$.
\label{lemma:phase21icbcrbb}
\end{lemma}
\begin{myproof}
 This follows trivially from the properties of binary consensus which guarantee that all honest
 nodes will eventually decide the same value. Since all honest nodes eventually decide the same value $b_i \in \{0,1\}$
 for binary consensus instance $i$, then, all honest nodes will eventually set $\mathfrak{B}[i]=b_i$, $\forall i \in [1,n]$.
 Thus, all honest nodes eventually agree on the same vector $\mathfrak{B}$.
\end{myproof}

\begin{table}[ht!]
\centering
\begin{tabular}{ | >{\centering\arraybackslash}m{2.5cm} | >{\centering\arraybackslash}m{2.5cm} | >{\centering\arraybackslash}m{2.5cm} |}
  \hline
  \multirow{2}{*}{Algorithmic Step} & \multicolumn{2}{c|}{Upper time bound of each clock} \\
  \cline{2-3}
  & $Clock$ & $Clock[n_i]$ \\ 
  \hline
  Honest nodes are initialized & \cellcolor{gray} 0 & $\Delta$ \\
  \hline
  Honest nodes multicast \emph{c-send} messages & $t_{comp} + 2\Delta$ & \cellcolor{gray} $t_{comp} + \Delta$ \\
  \hline
  Honest nodes receive \emph{c-send} messages of honest nodes & \cellcolor{gray} $t_{comp} + 2\Delta + \delta$ & $t_{comp} + 3\Delta + \delta$ \\
  \hline
  Honest nodes unicast \emph{c-ready} messages & $(n+1)t_{comp} + 4\Delta + \delta$ & \cellcolor{gray} $(n+1)t_{comp} + 3\Delta + \delta$ \\
  \hline
  Honest nodes receive \emph{c-ready} messages  of honest nodes & \cellcolor{gray} $(n+1)t_{comp} + 4\Delta + 2\delta$ & $(n+1)t_{comp} + 5\Delta + 2\delta$\\
  \hline
  Honest nodes multicast \emph{c-final} messages & $(n+2)t_{comp} + 6\Delta + 2\delta$ & \cellcolor{gray} $(n+2)t_{comp} + 5\Delta + 2\delta$\\
  \hline
  Honest nodes receive \emph{c-final} messages of honest nodes & \cellcolor{gray} $(n+2)t_{comp} + 6\Delta + 3\delta$ & $(n+2)t_{comp} + 7\Delta + 3\delta$ \\
  \hline
\end{tabular}
\caption{Upper time bounds for $Clock$ and $Clock[n_i]$ at each algorithmic step. Grayed cells indicate the reference point (clock variable)
with respect to which upper time bounds are calculated.}
\label{table:icbcrbbclocks}
\end{table}

When all $n$ parallel instances of binary consensus have completed, honest nodes will begin
finalizing each slot of the result vector according to the algorithm presented in Figure~\ref{figure:resultvectorfinalize}.
Recall, there is a special case where the outcome of a
binary consensus instance $i$ is $b_i=1$, but the private value of (faulty) node $n_i$ was not $obtained$ by an honest node $n_k$, by the end
of the value dissemination phase. In this case, $n_k$ will invoke the $Retrieve\_Value()$ procedure, 
presented in Figure~\ref{figure:retrieve}, to retrieve the private value of $n_i$. This will cause $n_k$ to multicast a
\textit{(retrieve, $n_i$)} message. Upon the receipt of a \textit{(retrieve, $n_i$)} message, an honest node $n_j$ executes the
algorithm presented in Figure~\ref{figure:onretrieve}. Provided that the value $v_i$ of node $n_i$ was $obtained$ by $n_j$ by
the end of the value dissemination phase, $n_j$ will reply to $n_k$ with a $valid$ \textit{(retrieved\_value, $n_i$, $v_i$, $c_i$)}
message. We say that a \textit{(retrieved\_value, $n_i$, $v_i$, $c_i$)} message is $valid$ \emph{iff} $c_i$ is a set of
$\langle n_j,v_j \rangle_{\sigma_k}$ tuples with valid signatures, from $(n-t)$ distinct nodes that
support the claim that $v_i$ is the private value of $n_i$. When an honest node executing the $Retrieve\_Value()$ procedure receives
such a $valid$ message, we say that it $retrieves$ the value $v_i$ of node $n_i$.

\begin{figure}[ht!]
 \footnotesize
 \begin{algorithmic}
  \STATE \textbf{procedure} $Finalize\_V():$
  \FOR{$1 \leq i \leq n$}
    \IF{$\mathfrak{B}[i] = 0$}
      \STATE $V[i] = \perp$;
      \STATE $C[i] = \perp$;
    \ELSE
      \IF{$V[i] = \perp$}
	\STATE $Retrieve\_Value(n_i);$
      \ENDIF
    \ENDIF
  \ENDFOR
 \end{algorithmic}
 \caption{Procedure $Finalize\_V()$ is invoked by honest nodes when all $n$ instances of binary consensus
 have completed to finalize the result vector $V$.}
 \label{figure:resultvectorfinalize}
\end{figure}

\begin{figure}[ht!]
 \footnotesize
 \begin{algorithmic}
  \STATE \textbf{procedure} $Retrieve\_Value(n_i):$
  \STATE send message \textit{(retrieve, $n_i$)} to all nodes;
  \STATE wait until the receipt of a $valid$ \textit{(retrieved\_value, $n_i$, $v_i$, $c_i$)} message;
  \STATE $V[i] = v_i$;
  \STATE $C[i] = c_i$;
 \end{algorithmic}
 \caption{Protocol $Retrieve\_Value()$ for retrieving $obtained$ values.}
 \label{figure:retrieve}
\end{figure}

\begin{figure}[ht!]
 \footnotesize
 \begin{algorithmic}
  \STATE Upon the receipt of a \textit{(retrieve, $n_i$)} from $n_k$ do
  \IF{$V[i] \neq \perp$}
   \STATE send a \textit{(retrieved\_value, $n_i$, $V[i]$, $C[i]$)} message to $n_k$;
  \ENDIF
 \end{algorithmic}
 \caption{Honest node behavior on the receipt of a \textit{(retrieve, $n_i$)} message.}
 \label{figure:onretrieve}
\end{figure}

\begin{lemma}
 Assume that the outcome of binary consensus instance $i$ is $b_i=1$. An honest node $n_j$
 executing the $Retrieve\_Value()$ procedure for input node $n_i$ will, eventually, complete the protocol. Additionally,
 the $retrieved$ value $v$ is the one broadcasted (using CB) by node $n_i$.
\label{lemma:retrieve1}
\end{lemma}
\begin{myproof}
%  According to the \emph{Validity} condition of binary consensus, if all non-faulty nodes have
%  the same initial value $b_i \in \{0,1\}$, then the agreed upon value by all non-faulty nodes will be $b_i$.
 Since the outcome of binary consensus instance $i$ is $b_i=1$, there is at least one non-faulty node that voted 1 for this
 binary consensus instance, and hence possesses the value broadcasted (using CB) by $n_i$. This honest node will always
 respond with a $valid$ \textit{retrieved\_value} message, as honest nodes comply with protocol rules.
 A malicious node, due to the $Integrity$ property of CB, cannot construct a $valid$ \textit{retrieved\_value} message for a
 value $v'$ which was not broadcasted by $n_i$. Thus, the behavior of malicous nodes is restricted to either replying with a
 $valid$ \textit{retrieved\_value} message, or not replying at all, since messages that are not $valid$ are rejected by honest
 nodes. Furthermore, if $n_i$ is faulty, the $Integrity$ property also guarantees that all honest nodes will $obtain$ at most
 one value, which, from the $Agreement$ property of CB, will be the same across all non-faulty nodes that actually $obtained$ it.
 Thus, an honest node executing the $Retrieve\_Value()$ procedure is guaranteed to eventually receive a $valid$ \textit{retrieved\_value}
 message and, thus, $retrieve$ the value $v$ broadcasted (using CB) by $n_i$.

\end{myproof}

\begin{theorem}
 The \emph{(IC,BC-RBB)} protocol satisfies the properties of IC.
\end{theorem}
\begin{myproof}

 \emph{Agreement}: Lemma~\ref{lemma:phase21icbcrbb} guarantees that all non-faulty nodes will eventually agree on the
 same vector of binary values $\mathfrak{B}$. If $\mathfrak{B}[i]=0$ for some $i \in [1,n]$, then all non-faulty nodes will
 deterministically set $V[i]=\perp$. If $\mathfrak{B}[i]=1$ we discriminate between two cases. First, if an honest node $obtained$
 a value $v_i$ of node $n_i$ by the end of the value dissemination phase, it takes no action since it has already
 set $V[i] = v_i$. Also, due to the $Agreement$ property of CB, all non-faulty nodes that $obtain$ a value for node $n_i$,
 $obtain$ the same value $v_i$. Second, if an honest node did not $obtain$ a value for node $n_i$ by the end of the
 value dissemination phase, it will, eventually, execute the $Recover\_Value()$ procedure. From Lemma~\ref{lemma:retrieve1}, it 
 follows that it will $retrieve$ the value $v_i$ that was broadcasted (using CB) by $n_i$ during the value dissemination phase and will,
 eventually, set $V[i] = v_i$. Thus, since every slot of the result vector $V$ will have the same value across all non-faulty nodes,
 it follows that all non-faulty nodes agree on the same vector of values $V$.
  
 \emph{Validity}:
 From Corollary~\ref{corollary:icbcrbb1}, it follows that the private value $v_i$ of an honest node $n_i$ has been $obtained$ by all non-faulty nodes by the
 end of the value dissemination phase. This will result in all non-faulty nodes to vote $b_i=1$ for binary consensus instance $i$, which
 will, eventually, lead all non-faulty nodes to set $\mathfrak{B}[i]=1$ due to the properties of binary consensus. Thus, all non-faulty nodes
 will, eventually, output $V[i]=v_i$.
  
 \emph{Termination}: From Lemma~\ref{lemma:phase21icbcrbb}, it follows that all non-faulty nodes will, eventually, execute procedure
 $Finalize\_V()$. We only deal with the $Retrieve\_Value()$ invocation, since all other steps require only local deterministic
 computation. From Lemma~\ref{lemma:retrieve1}, an honest node will always complete
 the $Retrieve\_Value()$ protocol. Thus, we have that all honest nodes will eventually complete the $Finalize\_V()$ procedure.
\end{myproof}

%% file: appendix.tex
\section{Message Complexities and Signature Operations}
\label{section:appendix2complexities}
In this section, we present the message complexity and signature operations for each algorithm presented in this 
paper. The term "signature operations" includes signature generations and verifications. We also assume graceful executions
of consensus, meaning that, when invoked, all consensus instances terminate without progressing to a second phase.

\subsection{Multicast (MU)}
A multicast primitive is comprised of a single step, during which, a node sends its
value to all other nodes, without any guarantee that every node will eventually deliver that
value.

The message complexity for invoking one instance of this multicast primitive equals to $n$ messages,
without requiring any signature operations.
\begin{itemize}
\item[]\emph{Message Complexity}: $n$
\item[]\emph{Signature Operations}: $-$
\end{itemize}

\subsection{Bracha's reliable broadcast (RBB)}
Bracha's reliable broadcast (\cite{bracha.async.consensus}) is comprised of three individual steps.
During the first step, the source node sends an \emph{initial} message to all other nodes ($n$ messages). During the second step,
each node that receives an \emph{initial} message, sends an \emph{echo} message to the rest nodes ($n^2$ messages). Finally, during
the third step, each node that receives an \emph{echo} message, sends a \emph{ready} message to all other nodes ($n^2$ messages).

The message complexity for invoking one instance of Bracha's reliable broadcast equals to $ 2n^2 + n $ messages, without requiring any signature operations.
\begin{itemize}
\item[]\emph{Message Complexity}: $2n^2 + n$
\item[]\emph{Signature Operations}: $-$
\end{itemize}

\subsection{Consistent Broadcast (CB)}
As presented in Figure~\ref{fig:cb}, Consistent Broadcast (\cite{cachin2001secure}) is comprised of three steps, each one with linear complexity.
During the first step, the source node sends a \emph{c\textrm{-}send} message to all other nodes ($n$ messages). During the second step,
each node that receives a \emph{c\textrm{-}send} message, sends back to the sender a signed \emph{c\textrm{-}ready} message ($n$ messages).
Finally, during the third step, the sender sends a \emph{c\textrm{-}final} message to the rest nodes of the system ($n$ messages). 

In addition, during the second step, each node generates one signature, before echoing the value back to the sender ($n$ signature generations).
Furthermore, during the third step, the sender verifies the signature of every received \emph{c\textrm{-}final} message
($n$ signature verifications). Finally, each node that receives a \emph{c\textrm{-}final} message, must verify its set of
signatures ($n^2$ signature verifications).

The message complexity for invoking one instance of Consistent Broadcast is $3n$ with $n^2 + 2n$ signature operations.
\begin{itemize}
\item[]\emph{Message Complexity}: $3n$
\item[]\emph{Signature Operations}: $n^2 + 2n$
\end{itemize}

% \subsection{Reliable Broadcast with Signatures (RBS)}
% The Reliable Broadcast with signatures is based on Consistent broadcast. More specifically, we introduce an additional step of communication 
% where all nodes send to all other nodes a \emph{final} message (Algorithm~\ref{algorithm:rbs}). This additional step adds to the previous message
% complexity $n^2$ extra messages.
% 
% During \emph{step 5}, nodes do not need to re-validate the signatures of each newly received \emph{final} message, because they have already validated
% that set of signatures, before sending their own \emph{final} message. As a result, this phase does not require any additional signature operations.
% 
% The message complexity for invoking one instance of Consistent Broadcast equals to $n^2 + 3n$, along with $n^2 + 2n$ signature operations.
% \begin{itemize}
% \item[]\emph{Message Complexity}: $n^2 + 3n$
% \item[]\emph{Signature Operations}: $n^2 + 2n$
% \end{itemize}

\subsection{Bracha's Binary Consensus (BC)}
The invocation of one instance of Bracha's binary consensus will trigger three instances of Bracha's reliable broadcast, without requiring any additional signature operations. The 
message complexity can be calculated by multiplying the complexity of reliable broadcast by a factor of three. This message complexity corresponds 
to the messages provoked by one node that participates in the consensus. To derive the entire message complexity of the algorithm we must also multiply the
previous result with a factor of \emph{n}. 

The message complexity for invoking one instance of Binary consensus (\textbf{BC-RBB}) equals to $n * 3 * RBB = n * 3 * (2n^2 + n) = 6n^3 + 3n^2$, since all nodes will broadcast their value.
For this algorithm, no signature operations are required.
\begin{itemize}
\item[]\emph{Message Complexity}: $6n^3 + 3n^2$
\item[]\emph{Signature Operations}: $-$
\end{itemize}

% By using the reliable broadcast with signatures (\textbf{BC-RBS}), the message complexity for invoking one instance of Binary consensus equals to $n * 3 * RBS = n * 3 * (n^2 + 3n) = 3n^3 + 9n^2$, since
% all nodes will broadcast their values. The total complexity of signature operations equals to $ 3 * n * RBS = 3 * n * (n^2 + 2n) = 3n^3 + 6n^2$.
% \begin{itemize}
% \item[]\emph{Message Complexity}: $3n^3 + 9n^2$
% \item[]\emph{Signature Operations}: $3n^3+ 6n^2$
% \end{itemize}

\subsection{Multi-valued Consensus (MC)}
To implement multi-valued consensus (\cite{correia.multivalue.consensus}), we use Bracha's reliable broadcast and Bracha's consensus algorithm (\cite{bracha.async.consensus}).
An instance of the multi-valued consensus protocol utilizes two instances of reliable broadcast and one instance of binary consensus.

The message complexity for invoking one instance of multi-valued consensus (\textbf{MC-RBB}) equals to $ n*2*RBB + (\textbf{BC-RBB}) = n*2*(2n^2 + n) + 6n^3 + 3n^2 = 10n^3 + 5n^2 $, without requiring any signature operations.
\begin{itemize}
\item[]\emph{Message Complexity}: $10n^3 + 5n^2$
\item[]\emph{Signature Operations}: $-$
\end{itemize}

% By using the reliable broadcast with signatures (RBS), the message complexity for invoking one instance of multi-valued consensus (\textbf{MC-RBS}) equals to 
% $ n*2*RBS + (\textbf{BC-RBS}) = n*2*(n^2 + 3n) + 3n^3 + 9n^2 = 5n^3 +  15n^2 $, along with $n*2*RBS + (\textbf{BC-RBS}) =  n*2*(n^2 + 2n) + 3n^3+ 6n^2 = 5n^3 + 10n^2$ signature operations.
% \begin{itemize}
% \item[]\emph{Message Complexity}: $5n^3+ 15n^2$
% \item[]\emph{Signature Operations}: $5n^3+ 10n^2$
% \end{itemize}

\subsection{Interactive Consistency BC-RBB}
The BC-RBB configuration is comprised of Consistent Broadcast and Bracha's binary consensus. The message complexity for invoking one instance of
BC-RBB equals to $ n*( CB + (\textbf{BC-RBB})) = n* ( 3n + (6n^3 + 3n^2) ) = 6n^4 + 3n^3 + 3n^2 $, along with $ n*( CB + (\textbf{BC-RBB})) = n*(n^2 + 2n) = n^3 + 2n^2 $ signature operations.
\begin{itemize}
\item[]\emph{Message Complexity}: $6n^4 + 3n^3 + 3n^2$
\item[]\emph{Signature Operations}: $n^3 + 2n^2$
\end{itemize}

% \subsection{Interactive Consistency BC-RBS}
% The BC-RBS configuration is comprised of Consistent Broadcast and Bracha's binary consensus, in which we substitute Bracha's reliable broadcast (RBB) with 
% Reliable broadcast using digital signatures (RBS).
% 
% The message complexity for invoking one instance of BC-RBS equals to $ n* (CB + (\textbf{BC-RBS})) = n*( 3n + (3n^3 + 9n^2)) = 3n^4 + 9n^3 + 3n^2$, along with 
% $ n* (CB + (\textbf{BC-RBS})) = n* (n^2 + 2n + (3n^3+ 6n^2)) = 3n^4 + 7n^3 + 2n^2$ signature operations.
% \begin{itemize}
% \item[]\emph{Message Complexity}: $3n^4 + 9n^3 + 3n^2$
% \item[]\emph{Signature Operations}: $3n^4 + 7n^3 + 2n^2 $
% \end{itemize}

\subsection{Interactive Consistency MC-RBB}
The MC-RBB configuration is comprised of a simple multicast primitive and multi-valued consensus. Each multicast primitive and multi-valued consensus will be invoked \emph{n} times, equal to the number
of nodes in the system.

The message complexity for invoking one instance of MC-RBB equals to $ n* (MU + (\textbf{MC-RBB})) = n*(n + (10n^3 + 5n^2)) = 10n^4 + 5n^3 + n^2$ without any signature operations.
\begin{itemize}
\item[]\emph{Message Complexity}: $10n^4 + 5n^3 + n^2$
\item[]\emph{Signature Operations}: $-$
\end{itemize}

% \subsection{Interactive Consistency MC-RBS}
% The MC-RBS configuration is comprised of a simple multicast primitive and multi-valued consensus using reliable broadcast with digital signatures (RBS). The configuration is invoked \emph{n} times, equal to the number
% of nodes in the system.
% 
% The message complexity for invoking one instance of MC-RBS equals to $ n* (MU + (\textbf{MC-RBS})) = n* (n + (5n^3 +  15n^2)) = 5n^4 + 15n^3 + n^2$ with
% $ n* (MU + (\textbf{MC-RBS})) = n* (0 + (5n^3 + 10n^2)) = 5n^4 + 10n^3$ signature operations.
% \begin{itemize}
% \item[]\emph{Message Complexity}: $ 5n^4 + 15n^3 + n^2$
% \item[]\emph{Signature Operations}: $5n^4 + 10n^3$
% \end{itemize}